\documentclass[twocolumn,groupedaddress]{revtex4-1}

\usepackage{ulem}
\usepackage{dcolumn}
\usepackage{bm}
\pdfpageattr{/Group << /S /Transparency /I true /CS /DeviceRGB>>} 
\usepackage{amsmath}
\usepackage{amssymb}
\usepackage{amsfonts} 
\usepackage{wasysym}  
\usepackage{graphics}  
\usepackage{psfrag} 
\usepackage{graphicx} 
\usepackage{epsfig}    
\usepackage{rotating}
 \usepackage[latin1]{inputenc}
 \usepackage{color}
 \usepackage{rotating}
\usepackage{csquotes}
 \usepackage{multirow}
 \usepackage{nicefrac}
\usepackage[nooneline, tight,raggedright,FIGTOPCAP]{subfigure}
\usepackage[linkcolor=red,citecolor=blue,urlcolor=blue,colorlinks=true,breaklinks=true]{hyperref}
\usepackage{breakurl}

\usepackage{color}
\usepackage{epsfig}
\usepackage{comment}

\begin{document}

\title{Microphase separation in active filament systems is \\ maintained by cyclic dynamics of cluster size and order}

\author{Lorenz Huber}
\thanks{L.H. and T.K. contributed equally to this work.}
\author{Timo Kr\"uger}
\thanks{L.H. and T.K. contributed equally to this work.}
\author{Erwin Frey}
\email{frey@lmu.de}
\affiliation{Arnold Sommerfeld Center for Theoretical Physics (ASC) and Center for NanoScience (CeNS), Department of Physics, Ludwig-Maximilians-Universit\"at M\"unchen, Theresienstrasse 37, D-80333 M\"unchen, Germany}


\begin{abstract}
The onset of polar flocking in active matter is discontinuous, akin to gas-liquid phase transitions, except that the steady state exhibits microphase separation into polar clusters.
While these features have been observed in theoretical models and experiments, little is known about the underlying mesoscopic processes at the cluster level.  
Here we show that emergence and maintenance of polar order are governed by the interplay between the assembly and disassembly dynamics of clusters with varying size and degree of polar order.
Using agent-based simulations of propelled filaments in a parameter regime relevant for actomyosin motility assays, we monitor the temporal evolution of cluster statistics and the transport processes of filaments between clusters.
We find that, over a broad parameter range, the emergence of order is  determined by nucleation and growth of polar clusters, where the nucleation threshold depends not only on the cluster size but also on its polar moment.
Growth involves cluster self-replication, and polar order is established by cluster growth and fragmentation.
Maintenance of the microphase-separated, polar-ordered state results from a cyclic dynamics in cluster size and order, driven by an interplay between cluster nucleation, coagulation, fragmentation and evaporation of single filaments. 
These findings are  corroborated by a kinetic model for the cluster dynamics that includes these elementary cluster-level processes.
It consistently reproduces the cluster statistics as well as the cyclic turnover from disordered to ordered clusters and back.
Such cyclic kinetic processes could represent a general mechanism for the maintenance of order in active matter systems.
\end{abstract}

\maketitle

\section{Introduction}
\begin{figure}[b]
\centering
\includegraphics[width=1.\columnwidth]{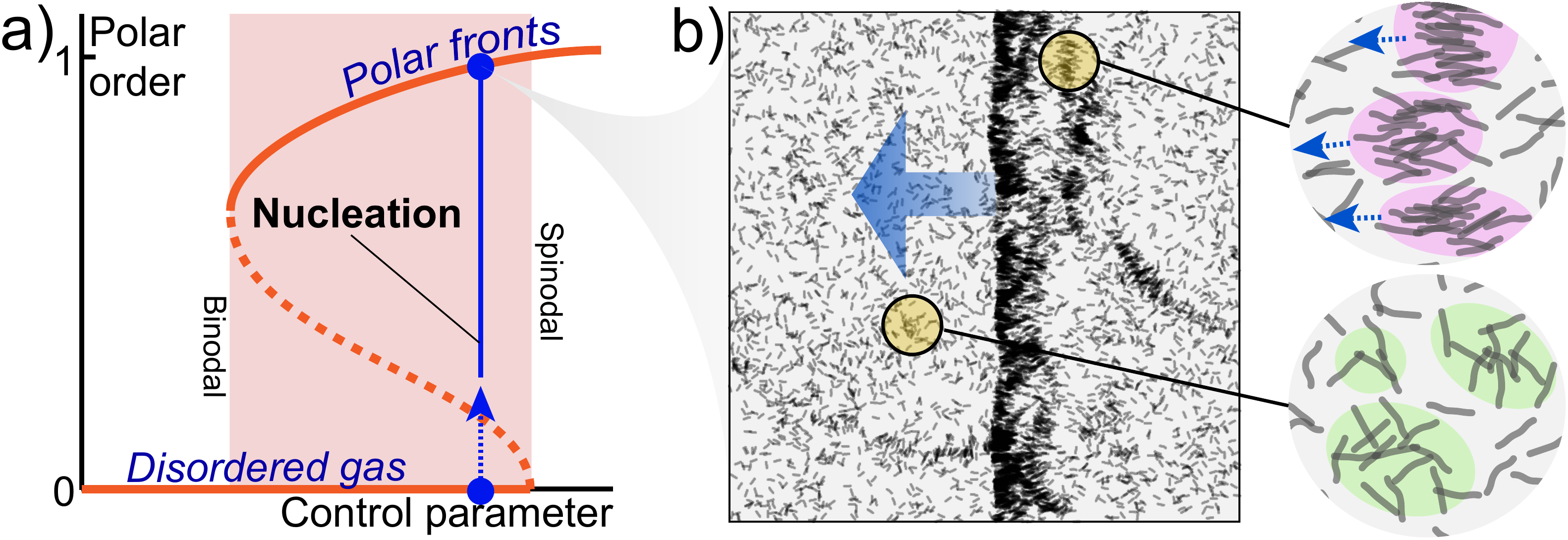}
\caption{a) Schematic of a typical bifurcation scenario for a flocking transition. Control parameters are, for example, particle density or interaction strength. Between binodal and spinodal, flocking is triggered by spontaneous nucleation events (blue line). (b) Illustration of clustering of active polymers in the polar, phase-separated state. Locally, both ordered (pink shading) and disordered (green shading) clusters are observed. 
}
\label{Fig::1}
\end{figure}

Polar flocking in active matter marks the onset of collective particle motion and has been observed in many experiments, ranging from biopolymer systems \cite{schaller_polar_2010, butt_myosin_2010, hussain_spatiotemporal_2013, suzuki_polar_2015,suzuki_emergence_2017} to colloids \cite{bricard_emergence_2013, kaiser_flocking_2017} and discs \cite{deseigne_collective_2010, deseigne_vibrated_2012, weber_long-range_2013, lam_self-propelled_2015}, as well as in theoretical studies using hydrodynamic descriptions \cite{bertin_boltzmann_2006,bertin_2009, mishra_fluctuations_2010,ihle_2011,gopinath_dynamical_2012,farrell_pattern_2012,marchetti_hydrodynamics_2013,grosmann_self-propelled_2013,ihle_2013,peshkov_boltzmann-ginzburg-landau_2014,caussin_emergent_2014,solon_pattern_2015-1} and particle based simulations \cite{gregoire_onset_2004, chate_collective_2008, solon_revisiting_2013, solon_phase_2015}. The associated nonequilibrium phase transition is in general discontinuous~\cite{gregoire_moving_2003, gregoire_onset_2004, bertin_boltzmann_2006,chate_collective_2008} and exhibits a subcritical parameter regime of polar patterns~\cite{solon_revisiting_2013, caussin_emergent_2014, thuroff_numerical_2014, solon_phase_2015, suzuki_polar_2015, morin_flowing_2018}, as illustrated in Fig.~\ref{Fig::1}(a).
While some aspects of flocking are akin to phase separation in thermal equilibrium systems~\cite{solon_flocking_2015, solon_phase_2015}, there are also marked differences.
In particular, both agent-based simulations and experiments have shown that active filament systems exhibit \textit{microphase separation} into dense polar-ordered regions and dilute disordered regions~\cite{chate_collective_2008, schaller_polar_2010, solon_phase_2015,shi_self-propelled_2018-1}.
How these steady-state patterns depend on the macroscopic control parameters (e.g.\ particle density, noise, or interaction strength) is well described at the level of hydrodynamic theories~\cite{caussin_emergent_2014, solon_phase_2015, solon_pattern_2015-1}.
The basic fact that spontaneous nucleation of particle clusters is vital for the initial stages of flocking is also well established~\cite{weber_nucleation-induced_2012, gregoire_onset_2004, chate_collective_2008}.
However, the mechanisms underling the formation and maintenance of a macroscopically ordered phase, which shows microphase separation into polar ordered clusters and a disordered background, is still unclear.

In the present work, we show that an interplay between cluster assembly and disassembly governs the emergence of polar order and microphase separation.
We find that particles self-organize into a heterogeneous population of clusters with a characteristic distribution of sizes and degree of polar order. 
By analyzing the temporal evolution of clusters using agent-based simulations of {\bf w}eakly {\bf a}ligning {\bf s}elf-propelled {\bf p}olymers (WASPs)~\cite{huber_emergence_2018}, we show that polar order and microphase separation in the flocking state are maintained by a continuous exchange of mass between coexisting populations of ordered and disordered clusters. 
To rationalize the underlying mechanism, we introduce a kinetic model consisting of two distinct cluster species, disordered and polar ordered, and study the ensuing assembly-disassembly dynamics.
We find that the kinetic model shows the same cluster statistics, mass-exchange dynamics, and bifurcation scenario as the agent-based system, even though it contains no information on the spatial dynamics. 
This theory explains the presence of microphase separation in the ordered state in terms of cyclic probability currents in a phase space spanned by cluster size and order.

\section{Results}
\subsection{Simulation setup and observables}
We consider agent-based simulations of a system with $M$ polymer filaments of fixed length $L$ on a two-dimensional substrate with periodic boundary conditions; for details see Ref.~\cite{huber_emergence_2018} and Appendix~\ref{sup:wasp_simulation_model}.
Motivated by experiments using \textit{in vitro} assays of gliding polymers \cite{sheetz_atp-dependent_1984, schaller_polar_2010, butt_myosin_2010, sanchez_spontaneous_2012, sumino_large-scale_2012, hussain_spatiotemporal_2013, suzuki_polar_2015,suzuki_emergence_2017}, each filament is assumed to consist of a head that performs a persistent random walk with persistence length $L_p$ and constant speed $v$, and a tail that passively follows it.
Interactions between filaments are assumed to be weak and dominated by aligning interactions \cite{suzuki_polar_2015, huber_emergence_2018}: 
upon local contact with adjacent filament contours, a polar and a nematic torque  proportional to $\varphi_p\cos \theta$ and $\varphi_n\cos 2 \theta$, respectively ($\theta$ being the impact angle), are exerted on the filament head.
These active filament systems were shown to reproduce local collision statistics and collective phenomena---polar and nematic patterns---on large scales ($M \,{=}\, \mathcal{O}(10^6)$)~\cite{huber_emergence_2018}, with filament density $\rho$ and relative alignment strength $\alpha \,{=}\, \varphi_n/\varphi_p$ as experimentally motivated control parameters. 
Here, we focus on the formation of large polar fronts as illustrated in Fig.~\ref{Fig::1}(b). 
In the flocking state, one observes that filaments are locally organized into clusters of different sizes and, on closer inspection, also of different degree of internal ordering [Fig.~\ref{Fig::1}(b)]: 
filament clusters in a polar front are highly ordered flocks while clusters elsewhere are much less structured. 

To investigate the role of clusters of different sizes and order in the emergence and maintenance of order in a system of WASPs, we monitor the size and degree of order of each filament cluster.
We decompose the system of filaments, $\{f_j \}$ with $j \,{\in}\, \{ 1, 2, \ldots, M \}$, into a set of clusters $\{ c_\alpha \}$: 
filaments are assumed to belong to a specific cluster $c_\alpha$ if they lie closer to filaments in that cluster than a cutoff distance $\gamma$ with $\gamma \,{\ll}\, L$, as described in more detail in Appendix~\ref{sup:cluster_polar_order_and_other}. 
Every cluster can be assigned a \textit{cluster size}, the number $k$ of filaments, and a \textit{cluster polar order}, $p_k \,{:=}\, \frac{1}{k}|\sum_{j=1}^k \exp(i\theta_j)|$. 
In the following it will turn out to be useful to also define the \textit{polar moment of a cluster}, $S_k \,{=}\, k \, p_k $, which measures the effective number of ordered filaments within a cluster.
Since even clusters made up of filaments with randomly chosen orientations have on average a nonzero polar order $\Delta_k\,{=}\, (7+\tfrac{1}{k})/(8\sqrt{k})+\mathcal{O}(k^{-5/2})$ [Appendix~\ref{sup:cluster_polar_order_and_other}], we define the net polar order of a cluster by $\pi_k \,{:=}\, p_k \,{-}\, \Delta_k$. 
Hence, the global polar order of the clusters   is given by an average of the net polar order $\pi_k$ weighted by the respective cluster size: 
$\Omega_p \,{:=}\,  \tfrac{1}{M} \sum_{\{c\}}\pi_k^{(c)} k^{(c)}$ (\textit{cluster polar order parameter}).
In addition to this system-level quantity, we also record the \textit{full statistics of cluster size and order}, $\Psi(k,p)$.
We choose a normalization such that the marginalized distribution $\psi(k) \,{=}\, \int_0^1\mathrm{d}p \,  \Psi(k,p)$ satisfies $\sum_{k=1}^M k \, \psi(k) \,{=}\, 1$.
This choice means that in a given realization (simulation run) $\psi(k) \,{=}\, n(k)/M$ where $n(k)$ is the number of clusters of size $k$;
hence, $\phi(k) \,{=}\, k \, \psi (k)$ gives the fraction of filaments contained in all clusters of size $k$. 
In the following we will refer to $\psi(k)$ as the cluster-size distribution.

\begin{figure}[!b]
\centering
\includegraphics[width=0.93\columnwidth]{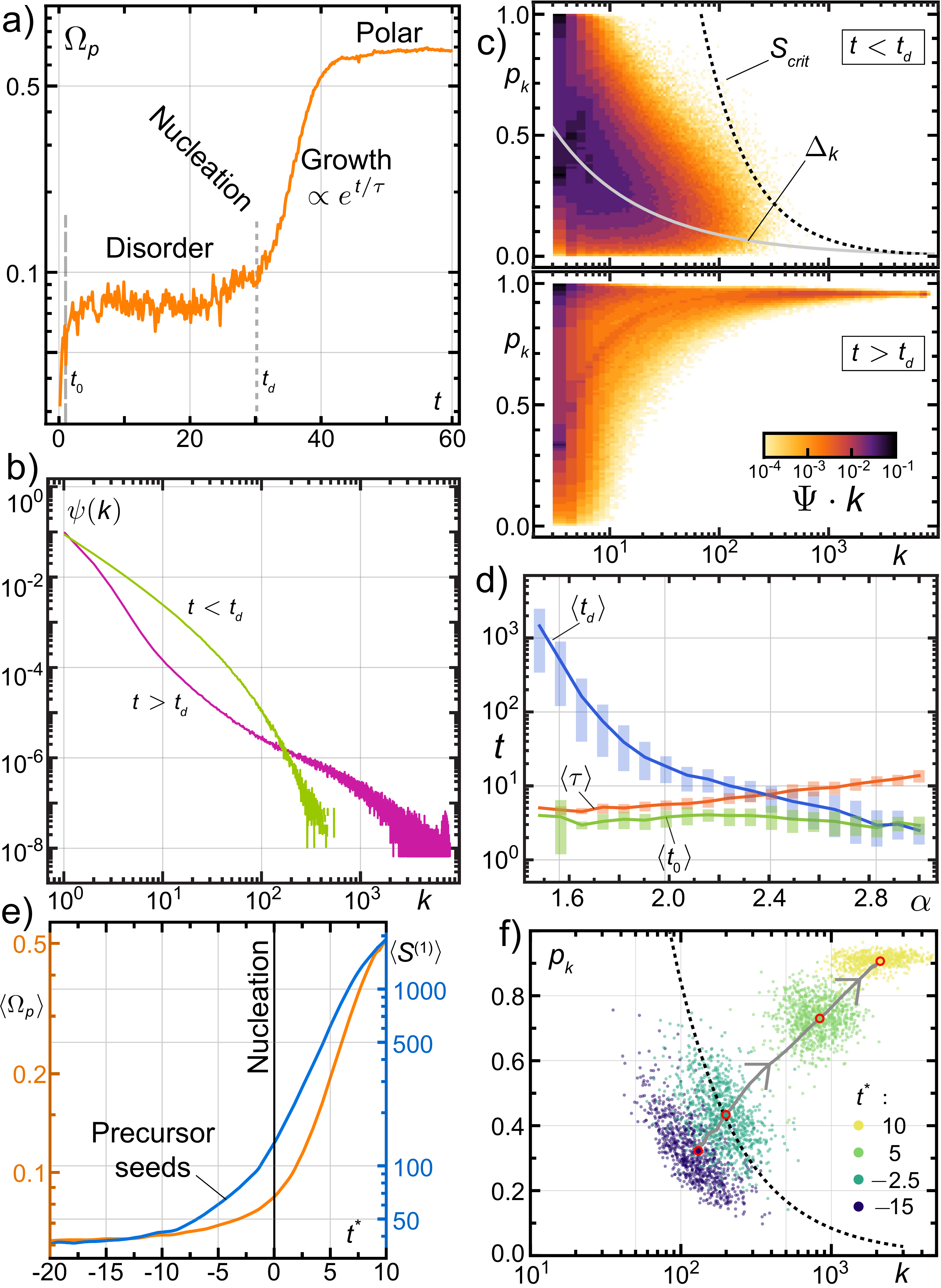}
\caption{
a) Time evolution of the cluster polar order parameter $\Omega_p$. We use units where time is expressed in terms of the longest single-particle correlation time $\tau_p \,{=}\, L_p/v$, i.e. the time over which the filament trajectories are approximately straight. The initial time scale $t_0$ and the nucleation time $t_d$ are marked by long-dashed and short-dashed lines, respectively.
b) Cluster size distribution, $\psi(k)$, in the disordered regime ($t \,{<}\, t_d$; green) and in the polar ordered steady state ($t \,{>}\, t_d$; purple). 
c) Heat plot (with color map shown in the graph) of the full statistics of cluster size and order, $k \cdot \Psi(k,p)$, plotted as a function of $k$ and $p$, in the disordered regime (upper panel) and in the polar ordered steady state (lower panel). The gray solid line depicts $\Delta_k$, and the dashed line indicates the estimated nucleation threshold $S_\text{crit} \,{=}\,  p_c k \approx 66$ (see discussion later). 
d) Characteristic time scales $t_0$, $t_d$, and $\tau$ as a function of $\alpha$. Solid lines denote average values, and error bars represent the $15$th, and $85$th percentiles taken over $100$ realizations for each $\alpha$.
e) Time evolution of $\langle S^{(1)}\rangle$ (blue line) and $\langle \Omega_p\rangle$ (orange line), as a function of $t^*\,{=}\,t \,{-\,} t_d$, averaged over $892$ independent realizations. 
f) Scatter plot for the size $k$ and order $p$ of the cluster corresponding to the largest cluster $S^{(1)}$ for $892$ independent realizations. The probability clouds at different times $t^*$ are indicated in different colors in the graph. As time progresses the cloud of points follows the trajectory indicated by the gray solid line, which depicts the average path of $\langle S^{(1)}\rangle$ in $k$-$p$ space. The red open circles mark the average $\langle S^{(1)}\rangle$ at the indicated timepoints. The dashed line indicates $S_\text{crit} \approx 84$. In panels a-c we used $\alpha \,{=}\, 2$, and in panels e-f a value of $\alpha \,{=}\, 1.67$. 
}
\label{Fig::2}
\end{figure}

\subsection{Polar order emerges through a hierarchical process}
To begin with, we show representative simulation results for the agent-based system in order to illustrate the dynamic processes that lead to the emergence of polar order starting from random initial conditions (as specified in Appendix~\ref{sup:wasp_implementation_and_parameters}).
If not stated otherwise, we fixed the parameters $\varphi_p \,{=}\, 0.036$ and $\rho \,{=}\, 1.51/L^2$. 

Figure~\ref{Fig::2}(a) depicts the time evolution of the cluster polar order parameter $\Omega_p$ for $\alpha \,{=}\, 2$, where the WASPs exhibit the same collision statistics as observed for actin filaments in the actomyosin motility assay slightly above the previously reported onset of flocking~\cite{huber_emergence_2018}; for an illustration of the associated dynamic processes please refer to Movie S1~(Appendix \ref{sup:movie_captions}).
We observe that generically within a relatively short time $t_0$ the system develops some but still rather weak polar order of the clusters with $\Omega_p \,{\approx}\, 0.08$.
The system persists in this disordered state for an extended time period until at some time $t_d$ cluster polar order suddenly and significantly increases and then approaches a stationary plateau value $\Omega_p^* \,{\approx}\, 0.7$; 
this growth phase is well described by an exponential law with the growth time $\tau$ [Fig.~\ref{Fig::2}(a)]. 
Visual inspection of the agent-based simulations suggests that the onset of polar order at $t_d$ is marked by the nucleation of a sufficiently large and polar-ordered cluster which triggers a cascade of cluster assembly and disassembly processes leading to rapid exponential increase in polar order; cf. Movie~S1~(Appendix \ref{sup:movie_captions}).

These qualitative observations are supported and quantified by the measured  statistics of cluster size and oder $\Psi(k,p)$. 
In the quasi-stationary, disordered regime ($t \,{<}\, t_d$) the distribution of cluster sizes, $\psi(k)$, shows an exponential tail [Fig.~\ref{Fig::2}(b)], similar to that found in previous studies \cite{huepe_intermittency_2004, peruani_nonequilibrium_2006, zhang_collective_2010, peruani_cluster_2010, yang_swarm_2010, peruani_kinetic_2013, fily_freezing_2014, duman_collective_2018}.
Moreover, the full distribution of cluster size and order, $\Psi(k,p)$, is centered around $p \,{\sim}\, \Delta_k$, indicating that typical clusters are only marginally more ordered than randomly assembled clusters [Fig.~\ref{Fig::2}(b,c)]. 
In contrast, in the stationary, polar-ordered state ($t \,{>}\, t_d$), the distribution of cluster size is no longer exponential but exhibits a broad tail [Fig.~\ref{Fig::2}(b)], and from the full statistics we infer that typical clusters are highly ordered~[Fig.~\ref{Fig::2}(c)].

Our simulations show that the onset times $t_d$ of polar order are randomly distributed, suggesting that nucleation events are stochastic and require rare events that initiate the formation of clusters of sufficiently large size and order.
Figure~\ref{Fig::2}(d) shows the mean and the statistical variation of the characteristic time scales $t_0$, $t_d$, and $\tau$ in the parameter range $\alpha \,{\in}\,  [1.5, 3.0]$; how these times are measured is detailed in Appendix~\ref{sup:time_scale_analysis}.
While the onset time $t_d$ of polar order increases strongly with decreasing $\alpha$, it remains finite even far below the previously reported onset of order at $\alpha \,{\approx}\,  1.8$~\cite{huber_emergence_2018}.
The onset times were found to be exponentially distributed with a  coefficient of variation $\sqrt{\mathrm{Var}[t_d]}/\langle t_d\rangle \,{\approx}\, 1$, similar as in classical nucleation theory~\cite{sear_nucleation_2007, kalikmanov_nucleation_2013}; for a detailed discussion of the observed variance in the onset time $t_d$ please refer to Appendix~\ref{sup:variance_of_the_nucleation_time}.
With increasing $\alpha$, we find that the average onset time $\langle t_d\rangle$ decreases and eventually becomes comparable to the average values $\langle t_0\rangle$ and $\langle \tau\rangle$, suggesting that the system instantly begins to develop polar order.
For even larger $\alpha$, polar order emerges through a process akin to spinodal decomposition (see discussion below and Movie S2~(Appendix \ref{sup:movie_captions}), which shows the dynamics for $\alpha \,{=}\, 3$).

\subsection{Nucleation barrier is determined by polar moment} 
To further characterize the processes underlying formation and growth of polar clusters we monitored the time evolution of all filament clusters and rank-ordered them according to the magnitude of their respective polar moments: $S^{(1)}\,{\ge}\, S^{(2)} \,{\ge}\, S^{(3)}\ge \dots \,{\ge}\, S^{(n)}$. 
Figure~\ref{Fig::2}(e) compares the time evolution of the cluster polar order parameter $\Omega_p$ and the largest polar moment $S^{(1)}$, averaged  over $892$ independent realizations and aligned in relation to the respective (stochastic) onset times $t_d$. 
The observation that growth of the largest cluster starts (on average) prior to the onset of polar order suggests that  precursor seeds  initiate cluster nucleation and growth. 
What then are their characteristic features? 
 
The answer becomes evident upon inspection of the evolution of cluster size and polar order, shown in Fig.~\ref{Fig::2}(f) as a scatter plot for different time points indicated in the graph; cf. Movie S3~(Appendix \ref{sup:movie_captions}).
Initially, before the onset time $t_d$, the probability cloud is widely extended in $k{-}p$ space and its center of mass hardly moves.
As soon as the cloud crosses a line of constant polar moment [dashed hyperbolic curve in Fig.~\ref{Fig::2}(f)], which occurs at a time that roughly coincides with the onset time $t_d$, we observe qualitatively different dynamics;
we will quantify the precise location of this transition line below.
The cloud then begins to contract and shows a clear trend  toward large cluster sizes $k$ and higher polar order $p$, i.e.\ increasing polar moment $S$. 
From these observations we conclude that the polar moment $S$ is the key quantity which determines the nucleation threshold. 

\subsection{Nucleation and spontaneous emergence of polar order}
\begin{figure}[b]
\centering
\includegraphics[width=\columnwidth]{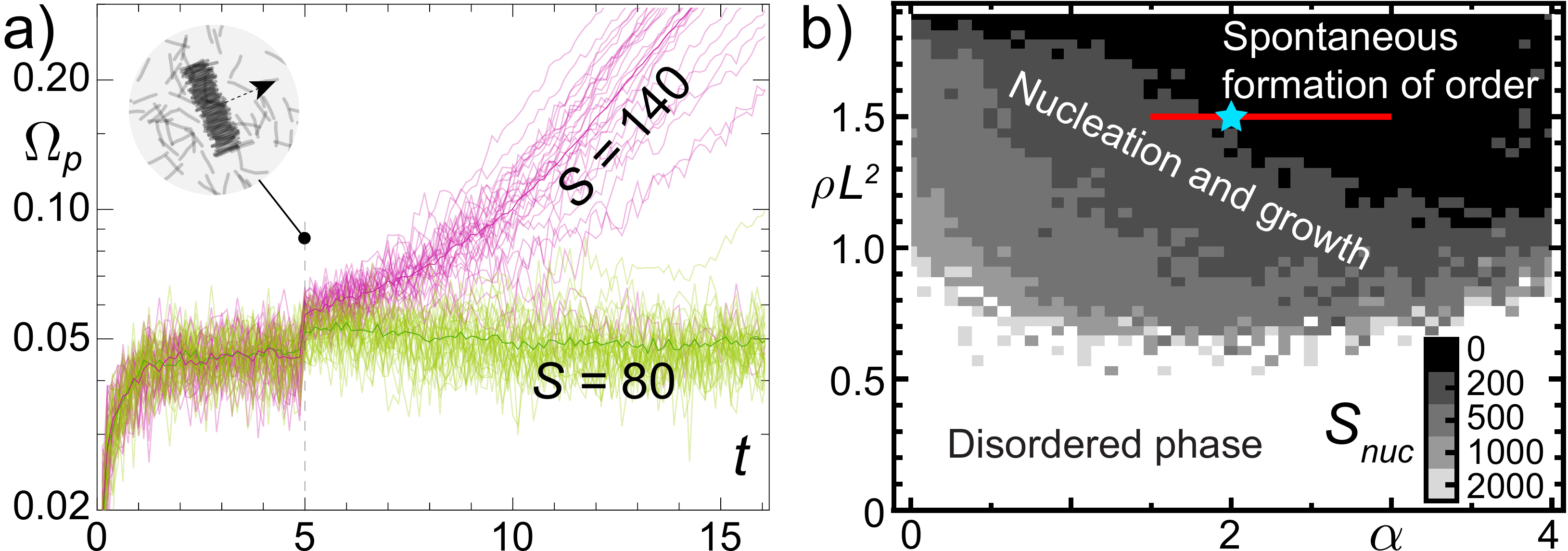}
\caption{
a) Time evolution of cluster polar order parameter $\Omega_p$ for disordered systems ($\alpha \,{=}\, 1.25$) perturbed by the addition of (fully polar) ordered cluster of polar moment $S$ at time $t \,{=}\, 5$ (green: $S\,{=}\, 80$; pink: $S \,{=}\, 140$). Thin lines correspond to single realizations, thick curves to the corresponding mean over all realizations. 
b) Phase diagram as a function of $\alpha$ and $\rho$. The regions shown in different shades of gray indicate regimes where the final system is polar-ordered with $\Omega_p^* \,{>}\, 0.2$. The gray scale corresponds to different values that are proxies for $S_\text{crit}$, as explained in the main text. The red line indicates the parameters used in Fig.~\ref{Fig::2}(d), and the blue star the parameters used in Fig.~\ref{Fig::2}(a-c).
}
\label{Fig::3}
\end{figure}
In order to determine the parameter regimes where polar order emerges either through a nucleation and growth process or spontaneously, we performed simulations over a wide range of densities, $\rho$, and relative alignment strengths, $\alpha$.
The black regime in Fig.~\ref{Fig::3}(b) indicates the parameter range, within which we observed onset times for polar order below $t_d \,{=}\, 50$.
We take this as a proxy for the regime where polar order builds up \textit{spontaneously}, cf. Movie S2~(Appendix \ref{sup:movie_captions}).
On the other hand, to determine the nucleation and growth regime and the respective threshold value of the polar moment (critical nucleus `size'), one would in principle need to monitor the time evolution of all clusters and wait for the spontaneous formation of a critical nucleus. 
While this is computationally feasible for parameter regimes where $t_d$ is reasonably small, it becomes practically impossible if $t_d$ is large, as is the case for small values of $\alpha$; c.f. Fig.~\ref{Fig::2}(d).
Therefore, we took a different approach and instead of waiting for a spontaneous nucleation event, we artificially inserted perfectly ordered ($p \,{=}\, 1$) clusters with different polar moments $S \,{=}\, k$ into a disordered system.
While clusters with $S \,{>}\, S_\text{crit}$ trigger a transition of the whole system towards a globally ordered state, the system remains disordered for smaller clusters, cf. an exemplary case in Fig.~\ref{Fig::3}(a). 
The different gray scales in Fig.~\ref{Fig::3}(b) show parameter regimes where nucleation and growth occurred in our simulations after insertion of a cluster of certain discrete size $S_\text{nuc}$ as indicated in the graph. 
These values correspond to proxies of $S_\text{crit}$ in the respective parameter regimes; 
see Appendix~\ref{sup:critical_polar_moment} for a more detailed analysis of $S_\text{crit}$.
For parameters where $t_d$ is small, we have explicitly checked that the critical value $S_\text{crit}$ obtained by artificially inserting a polar-ordered cluster and waiting for the spontaneous emergence of a critical nucleus agree quantitatively [Appendix~\ref{sup:critical_polar_moment_and_spont}]. 
On a qualitative level, this becomes evident from Movie S3~(Appendix \ref{sup:movie_captions}): The line given by $p(k)=S_\text{crit}/k$ defines a threshold curve in $k{-}p$ space, above which nucleation occurs, cf.\ also dashed curves in Fig.~\ref{Fig::2}(f). 
Moreover, upon comparing the course of nucleation for artificially triggered and spontaneous nucleation events in $k{-}p$ space, we found that very rapidly the emerging statistics for the largest cluster $S^{(1)}$ become indistinguishable from each other; see Fig.~\ref{Fig::A7} in Appendix~\ref{sup:course_of_nucleation_in}.
 
In summary, the phase diagram in Fig.~\ref{Fig::3}(b) exhibits two qualitatively distinct regimes. There is a regime where flocking is spontaneous akin to spinodal decomposition in liquid-gas systems, especially at high densities and large $\alpha$; cf. Movie S2~(Appendix \ref{sup:movie_captions}).
In addition, there is a broad range of parameters within which the transition to a polar ordered state proceeds by nucleation and growth.
In contrast to liquid-gas systems, the critical nucleus is not only characterized by a large enough size but also by a sufficiently high polar order, such that $S_\text{crit} \,{=}\, k \cdot p$.

\subsection{Coarsening and anti-coarsening} 
\begin{figure}[]
\centering
\includegraphics[width=1.\columnwidth]{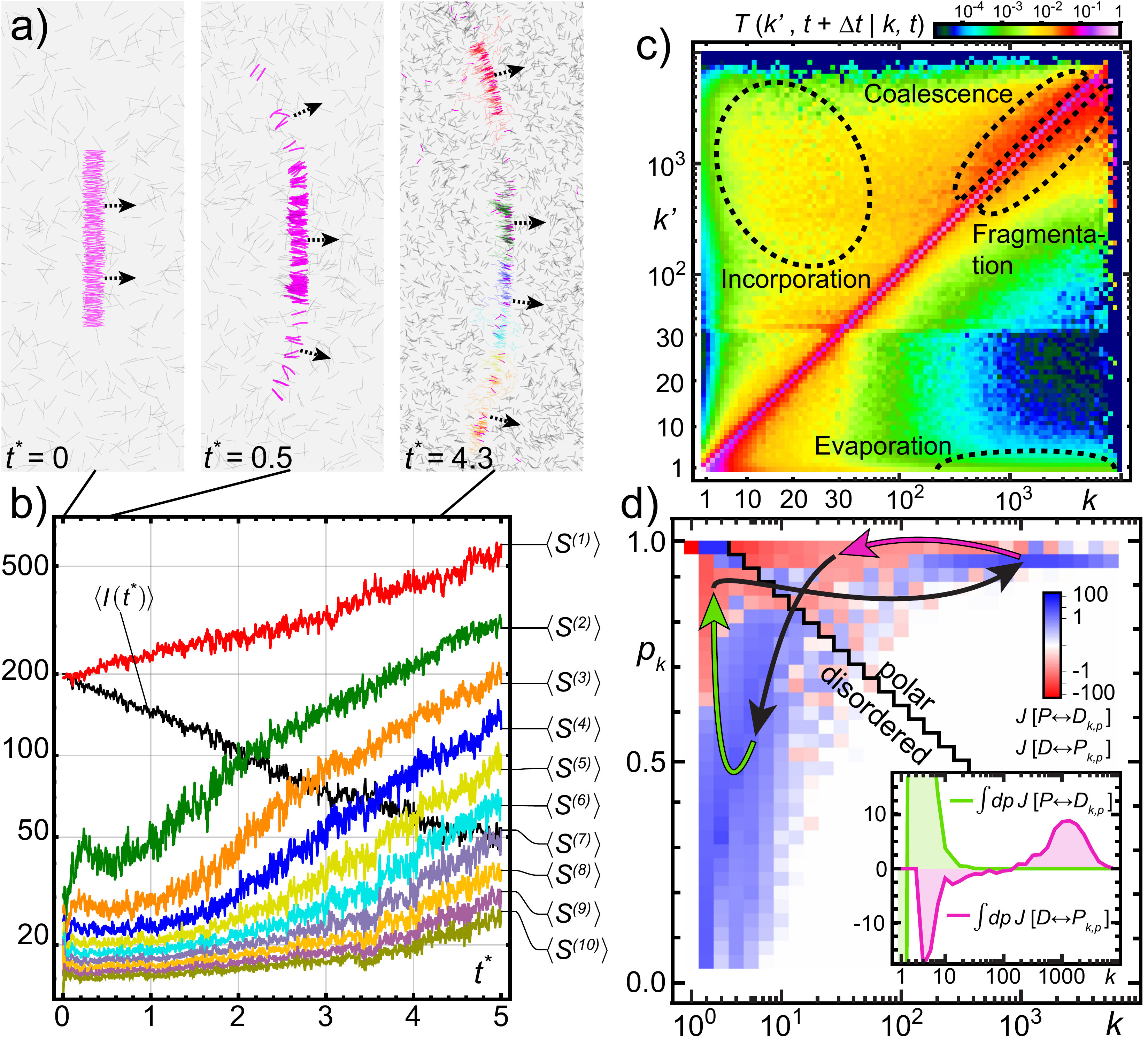}
\caption{
a) Snapshots of a perfectly ordered cluster added to a disordered system, taken at different times $t^*$ after insertion (at $t^* \,{=}\, 0$). Filaments that are part of the original cluster are shown in magenta. After growing for some time, the cluster eventually splits up into several distinct parts that can then grow on their own (shown in different colors). 
b) Time evolution of the clusters with the ten largest polar moments $S^{(i)}$, after an artificial nucleation seed of size $S_\text{seed} \,{=}\, 200$ was placed into the systems at $t^* \,{=}\, 0$, averaged over $30$ independent realizations. $I(t^*)$ specifies the temporal evolution of the amount of filaments which were originally part of the inserted cluster; cf. magenta filaments in panel a.  
c) Matrix of transition probabilities, $T(k',t \,{+}\, \Delta t | k, t)$, in color code as shown in the graph with $\Delta t\,{=}\, 0.0125$. As a guide to the eye, regions with dominant fragmentation or coalescence, incorporation or evaporation are encircled. 
d) Steady-state (in the polar-ordered phase) particle fluxes $J [D \,{\leftrightarrow}\, P_{k,p}]$ and $J [P  \,{\leftrightarrow}\, D_{k,p}]$ between ordered ($P$) and disordered ($D$) clusters in $k{-}p$ space as obtained from numerical simulations of WASPs. The black zig-zag line depicts the chosen partition of $k{-}p$ space into a disordered ($D$) and a polar ($P$) compartment. The arrows indicate the overall tendency in the flow between clusters of different size and polar order.
Inset: The fluxes $J [D  \,{\leftrightarrow}\, P_{k,p}]$ and $J [P  \,{\leftrightarrow}\, D_{k,p}]$ integrated over $p$ for comparison with Fig.~\ref{Fig::5}(e).
In all panels we used $\alpha \,{=}\, 1.67$. 
}
\label{Fig::4}
\end{figure}

Next, we wanted to gain further insight into the processes leading from the formation of a critical nucleus to the assembly of (moving) polar clusters and ultimately the polar-ordered, non-equilibrium steady state.
To this end, we artificially inserted seeds (fully ordered polar clusters) and observed their dynamics; 
for an illustration please refer to Fig.~\ref{Fig::4}(a) and Movie S4~(Appendix \ref{sup:movie_captions}). 
One observes that immediately after insertion the cluster begins to loose filaments. 
This loss is counteracted by a gain of filaments due to annexation of  disordered clusters (with low polar order) that lie in its pathway of motion. 
Only when the size of the seed is large enough, as discussed in the previous section, is this gain sufficient to overcome the filament loss such that the cluster grows.
These clusters, however, do not grow indefinitely, but eventually replicate by splitting up into several parts, which in turn  grow individually; frequently they also merge again.

These qualitative observations can be quantified in terms of the rank-ordered polar moments, whose averages sampled over $30$ realizations are shown for $S^{(1)}$ through $S^{(10)}$ in Fig.~\ref{Fig::4}(b). 
After artificial insertion of a seed cluster (here of size $S_\text{seed} \,{=}\, 200$), this seed forms the cluster with the largest polar moment $S^{(1)}$ which then grows exponentially, while one after another clusters with the next largest polar moment  follow suit. 
This sequential process corresponds to the continuous production of cluster fragments, which are created during splitting events and then grow by themselves.
The seed cluster spins off daughter clusters, as can be read off from the decline in the number of filaments $I$ that originally formed the seed cluster and are still part of the largest cluster $S^{(1)}$, cf. $I(t^*)$ in Fig.~\ref{Fig::4}(b).  

To further investigate the dynamics of clusters and the filament exchange between them, we tracked the fate of particles that were part of a cluster at time $t$ and recorded their status after some time $\Delta t$.
To this end, we define the transition probabilities $T(k',t \,{+}\, \Delta t | k, t)$ that quantify the likelihood that a filament which is part of a cluster of size $k$ at time $t$ will scatter into a cluster of size $k'$ at some later time $t \,{+}\, \Delta t$, normalized such that $\sum_{k'}T(k',t \,{+}\, \Delta t | k, t)\,{=}\, 1$; 
how $T$ is inferred from the simulation data is described in Appendix~\ref{sup:transiton_prob}.
For $\Delta t \,{\rightarrow}\, 0$, these transition probabilities become diagonal, $T(k',t | k, t)\,{=}\,\delta_{k k'}$, while for $\Delta t \,{\rightarrow}\, \infty$, as the events become statistically independent, one obtains $T(k',\infty | k, t)\,{=}\, k' \, \psi(k')$ [cf. Fig.~\ref{Fig::A9}(a,d) in Appendix~\ref{sup:evolution_of_the_cluster_distributions}]. 

Figure~\ref{Fig::4}(c) shows the matrix of these transition probabilities recorded for times $t$ in the stationary non-equilibrium steady state, and with the time increment chosen as $\Delta t\,{=}\, 0.0125$, a value corresponding to the time a filament takes to travel a distance comparable to its own contour length.
This choice gives each filament sufficient time to escape from its previous cluster, but multi-scattering events are still unlikely. 
The precise value of this time increment is not important [see Appendix~\ref{sup:transiton_prob}].
From Fig.~\ref{Fig::4}(c) we infer that, while most clusters remain stable during this time increment (diagonal), especially large polar clusters either frequently \textit{coalesce} or \textit{fragment} into similarly sized clusters (bright off-diagonal matrix elements in the upper right of Fig.~\ref{Fig::4}(c)), or \textit{evaporate} very small clusters or single filaments (bottom right matrix elements in Fig.~\ref{Fig::4}(c)). 
Clusters of smaller size, on the other hand, are frequently \textit{incorporated} into clusters of larger size  (upper left part of the matrix in Fig.~\ref{Fig::4}(c)).

Next, because of the qualitatively very different behaviour of strongly ordered and disordered clusters, we classified them into two broad classes: disordered ($D$) and polar ($P$) population. 
For that classification, we chose a heuristic division line in $k{-}p$ space [zig-zag line in Fig.~\ref{Fig::4}(d); cf. Appendix~\ref{sup:steady_state_flux_of_the_flocking}.
This is chosen such that in the quasi-stationary disordered regime [Fig.~\ref{Fig::2}(c), upper panel] most clusters would be classified as being disordered.
We monitored the net filament fluxes between these two populations in steady state. 
Specifically, we measured how many filaments transition per unit time between disordered/ordered clusters (of all sizes and degree of order) and ordered/disordered clusters of a given size and order, $J [D \,{\leftrightarrow}\, P_{k,p}]$ and $J [P \,{\leftrightarrow}\, D_{k,p}]$, respectively [Fig.~\ref{Fig::4}(d)]. 
These fluxes show that there is a cyclic flow of filaments between ordered and disordered clusters as indicated by the arrows in Fig.~\ref{Fig::4}(d): 
While large, ordered clusters show a net gain from disordered clusters, small ordered clusters lose to disordered clusters (black arrows). Since we are in steady state, i.e.\ particle numbers for each species must remain constant on average, there must also be net intra-species currents: (i) fragmentation of larger into smaller polar clusters (magenta arrow), and (ii) enhanced ordering of disordered clusters (green arrow).

Taken together, the above analysis of the agent-based simulations suggests that the following processes govern the emergence and maintenance of the stationary non-equilibrium steady state: 
In the quasi-stationary, disordered state the system consists of mostly disordered clusters with a wide distribution of sizes $k$ [Fig.~\ref{Fig::2}(b,c)]. 
Stochastically at time $t_d$, a critical nucleus (with polar moment of the order of $S_\text{crit}$) forms spontaneously, and subsequently grows exponentially by continuously incorporating more disordered clusters [Fig.~\ref{Fig::4}(b,c)]. 
By eventually splitting up [Fig.~\ref{Fig::4}(a-c)] due to orientational splay, polar clusters effectively self-replicate, which explains the exponential growth of the cluster polar order parameter $\Omega_p$ observed in Fig.~\ref{Fig::2}(a). 
In the final nonequilibrium steady state, there is a balance between different cluster-level kinetic processes: 
Growth of polar-ordered clusters through coagulation of polar-ordered clusters and incorporation of disordered filaments is balanced by splitting (fragmentation) of clusters as well as evaporation of smaller filament clusters back into the `pool' of disordered clusters [Fig.~\ref{Fig::4}(a,c)].
These processes drive the cyclical interconversion of the different types of cluster species, as indicated by the arrows in Fig.~\ref{Fig::4}(d).

\begin{figure}[b]
\centering
\includegraphics[width=\columnwidth]{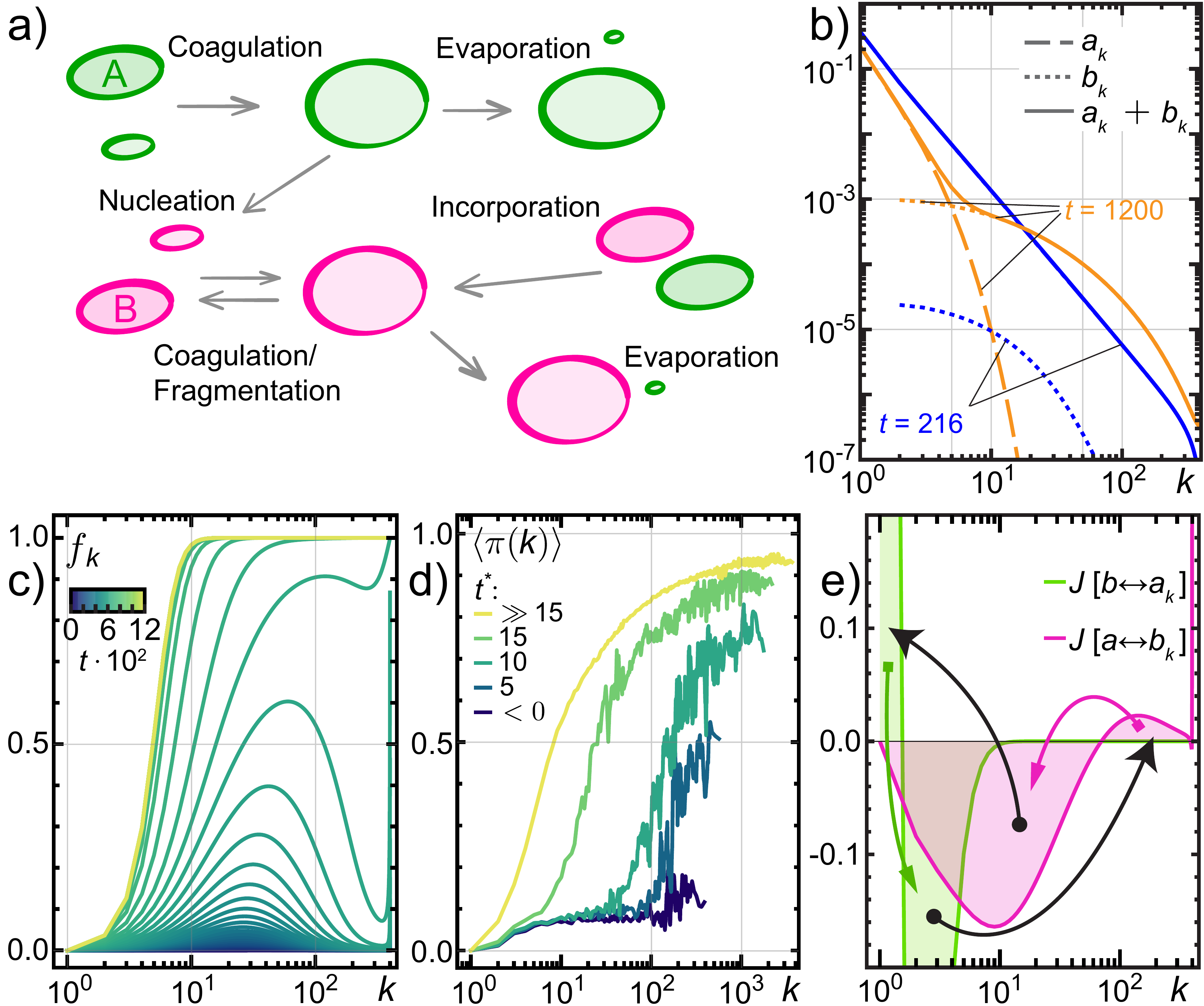}
\caption{
a) Illustration of the two-species kinetic model with a disordered cluster type A (green) and an ordered type B (purple) and the various cluster assembly and disassembly processes.
b) Time evolution of the cluster-size distributions, $a_k$ and $b_k$, of species A (long-dashed lines) and species B (short-dashed lines), respectively. The solid lines indicate the total distribution of cluster sizes, $n_k \,{=}\, a_k \,{+}\, b_k$, at two different times (blue at $t \,{=}\, 216$ and orange at $t \,{=}\, 1,200$).  
c) Time evolution of the relative fraction $f_k \,{=}\, b_k/n_k$. The color gradient depicts different times as quantified by the corresponding colour bar. 
d) Time evolution of the mean net cluster order $\langle \pi(k) \rangle_p$ during nucleation in the agent-based simulations. 
e) Steady-state particle fluxes $J [b {\leftrightarrow} a_k]$ and $J [a {\leftrightarrow} b_k]$ for both species as a function of cluster size $k$. Arrows: schematic depiction of inter-species (solid) and intra-species (solid colored) flux directions.
For the data shown for the kinetic model we have used the parameters: $M \,{=}\, 400$, $A \,{=}\, 800$, $v \,{=}\, \beta_0 \,{=}\, \lambda_0 \,{=}\, 1$, $\mu_0 \,{=}\, 0.025$, $\sigma_{aa} \,{=}\, 1.6$, $\sigma_{ab} \,{=}\, 0.2$, $\sigma_{bb} \,{=}\, 1$ and $\omega_0 \,{=}\, 10^{-4}$.
For the data shown in panel d) we used $\alpha \,{=}\, 1.67$.
}
\label{Fig::5}
\end{figure}

\subsection{Kinetic model for cluster assembly and disassembly} 
To determine whether these cluster assembly and disassembly processes constitute the essential mechanisms underlying the emergence and maintenance of the polar-ordered non-equilibrium steady state, we introduce a simple kinetic model; cf. Fig.~\ref{Fig::5}(a).
It reduces the dynamics of the spatially extended system to a set of kinetic processes for two competing types of cluster species, a \textit{disordered} type A and an \textit{ordered} type B, with respective cluster size distributions $a_k \,{=}\, (\mathbf{a})_k$ and $b_k \,{=}\, (\mathbf{b})_k$, where $\mathbf{x} \,{=}\, (x_1,x_2,...,x_M)$. 
The time evolution is assumed to be given by a set of coupled equations, $\partial_t {\mathbf{a}} \,{=}\, \mathbf{F}(\mathbf{a},\mathbf{b})$ and $\partial_t {\mathbf{b}} \,{=}\, \mathbf{G}(\mathbf{a},\mathbf{b})$ for the cluster size distributions, an approach frequently used to study coagulation and fragmentation dynamics in a broad class of systems~\cite{smoluchowski_versuch_1918, chandrasekhar_stochastic_1943, krapivsky_kinetic_2010}. The dynamics conserves the total number of  particles, $\sum_{k=1}^M k \, (a_k \,{+}\, b_k) \,{=}\, 1$.
Such kinetic models have successfully been used to describe the cluster statistics in a regime where polar order is absent~\cite{peruani_nonequilibrium_2006, peruani_cluster_2010, peruani_kinetic_2013}.
Our kinetic model extends these studies to include a second species B representing polar ordered clusters, and thereby enables us to study the assembly and disassembly processes leading to the emergence of polar order.

The set of nonlinear functions $\mathbf{F}$ and $\mathbf{G}$ --- for explicit forms see Appendix~\ref{sup:kinetic_model_equations} --- specify all the kinetic processes illustrated in Figure~\ref{Fig::5}(a): 
(i) For the disordered species A, cluster assembly occurs by \textit{coagulation} of smaller clusters of sizes $i$ and $j$ at a rate $\alpha_{ij} \,{:=}\, \sigma_{aa}  \,  X_{aa}(i,j) \, v /A$. Here $v$ is the cluster velocity, $A$ the area of the whole system, and $X_{aa}(i,j)$ a term dependent on the cluster sizes which characterizes the likelihood of cluster collisions. Since disordered clusters are approximately spherical in shape such that their diameter scales as $\sqrt{i}$, we take $X_{aa}(i,j) \, {=}\, \sqrt{i} \,{+} \sqrt{j}$. The parameter $\sigma_{aa}$ is an amplitude measuring the strength of the coagulation process of disordered clusters; in short: \textit{coagulation amplitude}.
(ii) Likewise, for the ordered species B, there is a \textit{coagulation} rate $\eta_{ij} \,{:=}\, \sigma_{bb} \, X_{bb}(i,j) \, v /A$. The elongated shape of ordered clusters suggests geometric factors that scale with their linear extension, $X_{bb}(i,j) \,{=}\, i \,{+}\, j$. Similar as above, the parameter $\sigma_{bb}$ designates the \textit{coagulation amplitude} for ordered clusters.
(iii) Ordered clusters of linear extension $i$ can \textit{incorporate} disordered ones of size $j$ at a rate $\gamma_{ij} \,{:=}\, \sigma_{ab} \, X_{ab}(i,j) \, v /A$, and thereby form a larger ordered cluster. The geometric factor is now assumed to be $X_{ab}(i,j) \,{=}\, i$, and $\sigma_{ab}$ is called the \textit{incorporation amplitude}. 
(iv) Cluster disassembly occurs via \textit{split-up} (\textit{fragmentation}) of ordered clusters at a constant rate $\mu_{ij} \,{=}\,\mu_0$, and \textit{evaporation} of single disordered particles from cluster species A and B at rates $\beta_i \,{:=}\, \beta_0 \, Y_a(i)$ and $\lambda_i  \,{:=}\,  \lambda_0 \, Y_b(i)$, respectively. The geometric factors read $Y_a(i) \,{=}\, \sqrt{i}$ and $Y_b(i)=1$, where the latter accounts for the observation that ordered waves evaporate particles mainly via its edges, i.e.\ there is no size dependence. 
(v) Finally, a disordered cluster may spontaneously \textit{transform} into an ordered cluster, at a rate $\omega_i \,{:=}\, \omega_0 \, Z(i)$ with $Z(i) \,{=}\, 1/(1+e^{-(i-m_c-1)/v_c})$; this event effectively represents the \textit{nucleation} of an ordered cluster. The sigmoidal shape accounts for the observation that nucleation only occurs above a certain threshold cluster size $m_c$. For specificity we choose $m_c \,{=}\, 100$ and $v_c \,{=}\, 10$ throughout our analysis. Variation of $m_c$ or $v_c$ results only in a shift in the onset of the transition to polar order, without any qualitative effects on the ordered state; cf. Appendix~\ref{sup:detailed_form_of_transformation_rate}.

The  kinetic model is not an exact representation of the kinetics observed in the agent-based model, but it emulates its core features.
First, global polar order in the system of WASPs is analogous to the mass fraction of the ordered species $\phi_b \,{=}\, \sum_k k \, b_k$ in the kinetic model. 
Second, while the parameter $\alpha$ quantifies the (relative) strength of the alignment interaction responsible for flocking of WASPs, the corresponding analogs in the kinetic model are the amplitudes $\sigma_{ab}$ and $\sigma_{bb}$ that quantify the strength of processes leading to an increase in polar order $\phi_b$.  
In the following, we describe the influence of these parameters on the size distributions $a_k$ and $b_k$.
For the coagulation amplitude $\sigma_{aa}$ of the disordered clusters we chose a fixed value of $\sigma_{aa} \,{=}\, 1.6$,  such that --- in the absence of an ordered species B --- the size distribution $a_k$ resembles the previously observed exponentially truncated power law~\cite{peruani_nonequilibrium_2006, peruani_cluster_2010, peruani_kinetic_2013};  cf. Fig.~\ref{Fig::2}(b).
We integrated the set of kinetic equations to find the time evolution of the distribution of cluster sizes, $\{a_k (t), b_k(t) \}$, using a simple Euler scheme, and starting from initial conditions where all particles were in clusters of size $k \,{=}\, 1$: $a_1(0) \,{=}\, 1$.
If not stated otherwise, we used the parameters specified in Fig.~\ref{Fig::5}. 

To begin with, we present the results for specific amplitudes: $\sigma_{ab} \,{=}\, 0.2$ and $\sigma_{bb} \,{=}\, 1$. 
In that case, the distribution of total cluster sizes, $n_k \,{:=}\, a_k+b_k$, changes with time from an exponentially truncated power-law form [blue solid line in Fig.~\ref{Fig::5}(b)] to a broad distribution with a distinct shoulder at intermediate $k$ [orange solid line in Fig.~\ref{Fig::5}(b)], similar to the results obtained for a system of WASPs [Fig.~\ref{Fig::2}(b)]. 
How polar order emerges is also quite comparable, as can be inferred from the time evolution of the fraction of ordered clusters,  $f_k \,{:=}\, b_k/n_k$, in the kinetic model [Fig.~\ref{Fig::5}(c)] and the mean net cluster order, $\langle \pi (k) \rangle_p \,{:=}\, \int_0^1 \mathrm{d}p \, \pi_k \Psi(k,p)$, in the agent-based simulations [Fig.~\ref{Fig::5}(d)]. 
In both instances, ordered clusters begin to proliferate at intermediate sizes $k$, followed by a broadening of the distribution towards smaller as well as larger cluster sizes. 

Next, as in the case of the agent-based model [cf. Fig.~\ref{Fig::4}(c,d)], we wanted to learn how the various kinetic processes operating within species and between ordered and disordered clusters balance to maintain a stationary polar-ordered state, where $\partial_t a_k \,{=}\, 0 \,{=}\, \partial_t b_k$.
For each species and each cluster size $k$, this requires a strict balance between inter-species and intra-species currents.
Moreover, note that there is also a global balance such that the total number of particles remains constant. 
Figure \ref{Fig::5}(e) shows the net inter-species currents $J [a {\leftrightarrow} b_k] $ (magenta) and $J [b {\leftrightarrow} a_k] $ (green) for the ordered and disordered species, respectively; intra-species currents are simply the opposite, e.g.\ for the ordered species: $J [b {\leftrightarrow} b_k] \,{=}\, {-} \, J [a {\leftrightarrow} b_k] $.
For the ordered clusters, $J[a {\leftrightarrow} b_k] \,{<}\, 0$ for a wide range of cluster sizes, indicating that there is an overall net loss of ordered clusters in favor of disordered clusters.   
A more detailed analysis shows that this is largely due to \textit{evaporation} of single disordered particles [see Appendix~\ref{sup:Dynamical_and_steady_state_properties}]. 
At large cluster sizes, there is a net gain ($J[a {\leftrightarrow} b_k] \,{>}\, 0$) in the number of ordered clusters, which can be attributed to the \textit{incorporation} of disordered clusters by ordered clusters.
The balance between intra-species and inter-species processes requires that there is a net flux from large to small ordered clusters, i.e.\ a surplus of cluster \textit{fragmentations} relative to cluster \textit{coagulation} events. 
This is phenomenologically similar to our findings in the agent-based simulations, cf. Fig.~\ref{Fig::4}(d).
There, we observed that large ordered clusters gain from disordered clusters, and small ordered clusters loose filaments to disordered clusters.
This implies that there must be an intra-species current within ordered clusters, presumably also mediated by splitting of large into smaller ordered clusters. 
For the disordered clusters, we observe a net gain ($J[b {\leftrightarrow} a_k] \,{>}\, 0$) of single disordered particles, which is due to \textit{evaporation} events from ordered clusters.
On the other hand, there is a net loss ($J[b {\leftrightarrow} a_k] \,{<}\, 0$) of disordered clusters at intermediate cluster sizes, which is due to \textit{incorporation} of disordered clusters into ordered clusters (and to smaller extent due to spontaneous transformation of disordered into ordered clusters). 
As the inter-species processes with ordered clusters create a surplus of single disordered particles, in steady state this must be balanced by a corresponding intra-species flux from small to large disordered clusters, which is facilitated by \textit{coagulation} processes of disordered clusters.

\begin{figure}
\centering
\includegraphics[width=\columnwidth]{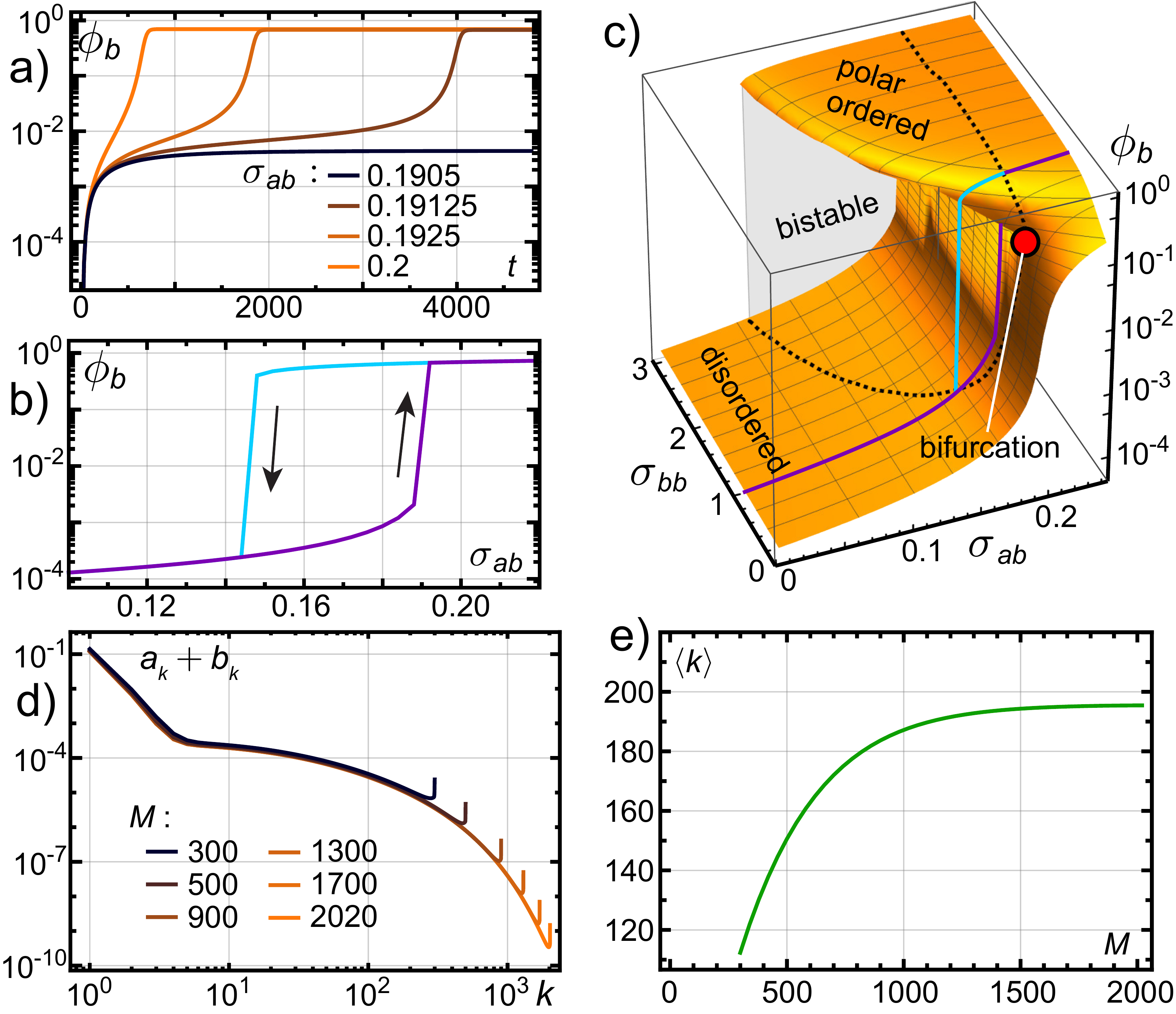}
\caption{ 
a) Evolution of the mass fraction $\phi_b$, for different values of $\sigma_{ab}$ ($\sigma_{bb} \,{=}\, 1$). 
b) Hysteresis of the stationary mass fractions $\phi_b$ as a function of $\sigma_{ab}$.
c) Bifurcation diagram of stationary mass fractions $\phi_b$ as a function of $\sigma_{ab}$ and $\sigma_{bb}$. The dashed lines mark the upper and lower boundaries of the bistable region, respectively. The coloured lines mark the position of the data shown in b). 
d) Stationary total cluster distribution $a_k+b_k$ as a function of the system size $M$.
e) Stationary mean cluster size $\langle k\rangle$ as a function of the system size $M$.
In panels d-e we used $\sigma_{aa}\,{=}\,1.4$, $\sigma_{ab}\,{=}\,0.2$, $\sigma_{bb}\,{=}\,0.8$, $\mu_0\,{=}\,0.01$ and $\omega_0 \,{=}\, 10^{-5}$.
}
\label{Fig::6}
\end{figure}

In order to determine the phase diagram and the nature of the corresponding phase transitions, we studied how the emergence of polar order in the kinetic model depends on the strength of the various processes.
We focused on the effects of \textit{coagulation} of ordered clusters and the \textit{incorporation} of disordered clusters into ordered clusters, varying the corresponding amplitudes $\sigma_{bb}$ and $\sigma_{ab}$, respectively.  
Figure~\ref{Fig::6}(a) shows the time evolution of the mass fraction $\phi_b$ of the ordered B species for various values of the incorporation amplitude $\sigma_{ab}$. 
Like the cluster polar order parameter $\Omega_p$ [Fig.~\ref{Fig::2}(a)] it exhibits a transient dwelling period before (exponentially fast) approaching the polar-ordered states.
Interestingly, the duration of this dwelling time seems to be very sensitive to changes in the overall incorporation rate $\sigma_{ab}$ [Fig.~\ref{Fig::6}(a)]. 
In addition, in accordance with our agent-based simulations  [Fig.~\ref{Fig::3}(b)] and as found in previous studies~\cite{gregoire_moving_2003, gregoire_onset_2004, bertin_boltzmann_2006, chate_collective_2008, solon_revisiting_2013, caussin_emergent_2014, thuroff_numerical_2014, solon_phase_2015, suzuki_polar_2015, morin_flowing_2018}, the order parameter $\phi_b$ shows a discontinuity and hysteresis as a function of a control parameter [Fig.~\ref{Fig::1}(a)], here the incorporation amplitude $\sigma_{ab}$ [Fig.~\ref{Fig::6}(b)].
Varying both $\sigma_{ab}$ and $\sigma_{bb}$, we obtain the bifurcation diagram (for the stationary state) shown in Fig.~\ref{Fig::6}(c); please refer to Appendix~\ref{sup:parameter_space_and_hysteresis} for a bifurcation diagram as a function of density $\rho_\text{kin}$ and $\sigma_{bb}$.
The effects of coagulation of ordered clusters and incorporation of disordered clusters by ordered clusters on the emergence of polar order are quite distinct.
While the amplitude of the incorporation processes ($\sigma_{ab}$) appears to regulate the transition from a disordered to a polar-ordered state, the amplitude of the coagulation processes of ordered clusters ($\sigma_{bb}$) affects the character of this phase transition.
For small $\sigma_{bb}$ (weak propensity for coagulation of ordered clusters), the transition is continuous, and becomes discontinuous only above a certain threshold value, with the ensuing bistable parameter regime broadening as $\sigma_{bb}$ increases further. 

Finally, we checked whether the kinetic model also exhibits microphase  separation, as observed in other models~\cite{chate_collective_2008, schaller_polar_2010, solon_phase_2015,shi_self-propelled_2018-1}.
To this end, we increased $M$ (adapting the area $A$ to keep the density constant) and recorded its influence upon the stationary total cluster distribution $n_k \,{=}\, a_k \,{+}\, b_k$, as well as the stationary mean cluster size $\langle k \rangle$ [Fig.~\ref{Fig::6}(d,e)]. 
Notably, both become independent of system size above a certain value of $M$.
We conclude that the polar phase of the kinetic model also exhibits arrested growth and hence microphase separation, like that observed in polar active systems~\cite{chate_collective_2008, schaller_polar_2010, solon_phase_2015,shi_self-propelled_2018-1}. 
This contrasts with the single-species model of Peruani et al. \cite{peruani_nonequilibrium_2006, peruani_cluster_2010, peruani_kinetic_2013} which exhibits a continuous order transition from a state with microscopic clusters towards a macrophase separated state. 

\section{Discussion}

An intriguing phenomenon in polar active matter is not only the emergence of polar ordered clusters, but also the fact that the ordered state exhibits microphase separation into dense, polar-ordered clusters and a gas-like disordered filament reservoir. 
Here we asked how the kinetic processes of cluster assembly and disassembly might reveal the underlying mechanism.
To answer this question we used a two-pronged approach based on agent-based simulations and a corresponding cluster-level kinetic theory. 
Our main conclusion is that microphase separation in polar active matter is a cyclic self-organizing process of particle clusters of different sizes and degrees of polar order rather that a halted coarsening process.

Using agent-based simulations we monitored the kinetic processes at both the particle and the cluster level and thereby determined the time evolution of the cluster statistics in terms of cluster size and degree of polar order.
Moreover, these simulations also allowed us to fully relate the mesoscopic cluster dynamics to the underlying microscopic dynamics of individual filaments.
Taken together, this yielded the following key insights:  
First, we find two qualitatively distinct parameter regimes, one where polar order emerges spontaneously and another which requires the formation of a nucleus and its subsequent growth.
Our simulations show that the nucleation barrier is not determined by either cluster size $k$ or cluster order $p_k$ alone, but by the polar moment $S_k \,{=}\, k \cdot p_k$.
Second, once a critical nucleus has formed, an intricate dynamics of cluster assembly and disassembly processes is triggered that leads to microphase separation between high-density, polar-ordered clusters and a low-density, disordered background. 
It entails the growth of clusters by the incorporation of disordered filaments, the breakup of larger into smaller sub-clusters and their subsequent growth (cluster self-replication), coalescence of clusters and evaporation of filaments from ordered clusters into the disordered background.
We have quantified these processes in terms of the probability currents between clusters of different size $k$ and degree of polar order $p$.
This analysis suggests that the dynamics that maintains a non-equilibrium steady state is a cyclic dynamics in $(k,p)$ phase space.

These results suggested that the dynamics of the active filament system can be understood in terms of kinetic processes at the mesoscopic level of clusters, i.e.\ by considering the assembly and disassembly of clusters with different size and degree of order.
To test this hypothesis we formulated a simple kinetic model that emulates the key processes identified in the agent-based simulations and analyzed the same or analogous observables.
The kinetic model shows the same phenomenology as the agent-based simulations, including similar probability flows in phase space and the same topology of the bifurcation diagram.
Most importantly, the kinetic model exhibits arrested growth and hence microphase separation. 
That opens a new perspective on this phenomenon: instead of focusing on a characterization of the spatio-temporal patterns we identify the relevant kinetic processes that govern the probability flow in phase space.

We propose that kinetic descriptions, similar to the one introduced here, might already capture the essential dynamics of other collective phenomena in active systems, such as nematic laning \cite{ginelli_large-scale_2010, peshkov_nonlinear_2012, ngo_large-scale_2014}, vortex formation \cite{sumino_large-scale_2012, loose_bacterial_2014, denk_active_2016} or coexisting types of order~\cite{huber_emergence_2018,gromann2019_particlefield,denk_pattern-induced_2020}.
In particular, the flow in a properly defined phase space might reveal, as we show here, the mechanisms that underlie the emergence and maintenance of the corresponding non-equilibrium steady states.

\begin{acknowledgements}
This work was funded by the Deutsche Forschungsgemeinschaft (DFG, German Research Foundation) -- Project-ID 111166240 --  Collaborative Research Center (SFB) 863 - Project B2.
\end{acknowledgements}

\appendix

\section{WASP simulations} \label{sup:wasp_simulations} 
In the following, we shortly discuss the implementation of the agent-based simulations of {\bf w}eakly-{\bf a}ligning {\bf s}elf-propelled {\bf p}olymers (WASP's). 
For a detailed description please refer to the Supplemental Material of Ref.~\cite{huber_emergence_2018}.

\subsection{WASP simulation model} \label{sup:wasp_simulation_model} 
\begin{figure}[b]
\centering
\includegraphics[width=0.8\columnwidth]{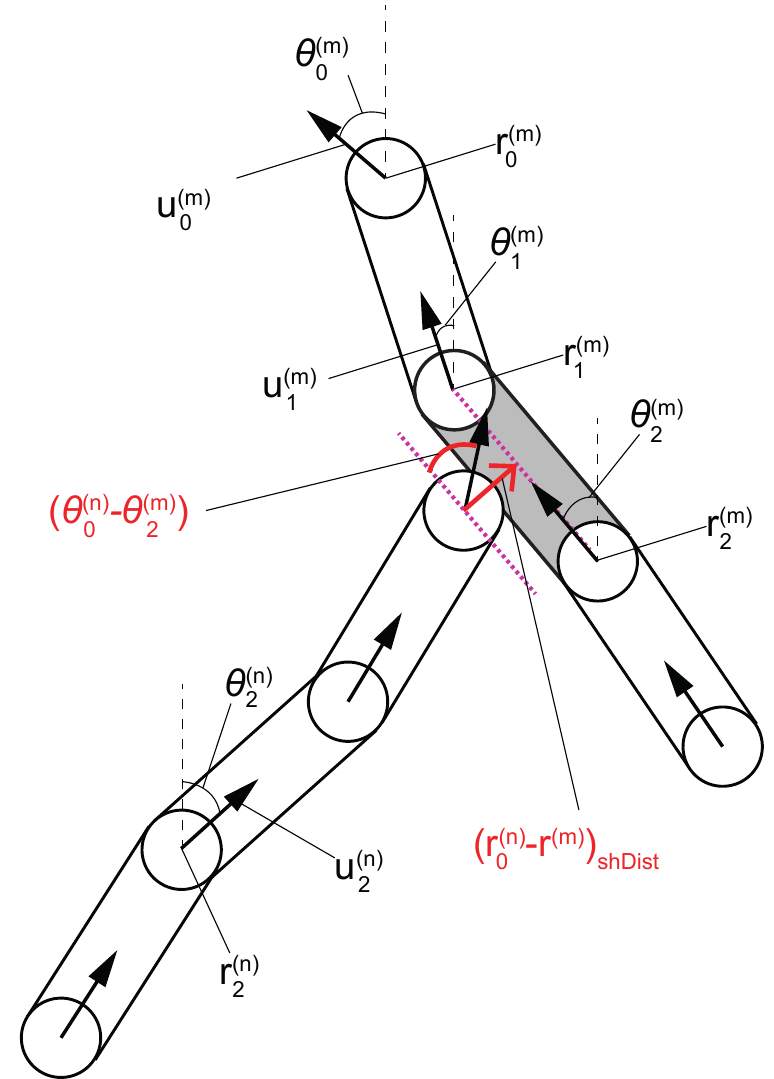}
\caption{
\textbf{Illustration of interactions in the filament model.} 
The head of a filament $n$ collides with the body (contour) of an adjacent filament $m$ between bead position $\mathbf{r}_1^{(m)}$ and $\mathbf{r}_2^{(m)}$. 
The \textit{impact angle} between the two filaments is given by $\Delta \theta_{nm} \,{:=}\, \theta_0^{(n)} \,{-}\,\theta_2^{(m)}$, where $\theta_0^{(n)}$ and $\theta_2^{(m)}$ denote the orientation of the head of the $n^\text{th}$ polymer and the orientation of the tangent to the body of the $m^\text{th}$ polymer where the collision happens. 
In the illustrated case the latter is given by the orientation of the $2^\text{nd}$ cylinder of the $m^\text{th}$ polymer (which in turn is given by the orientation of the normalized bond vector, $\mathbf{u}_2^{(m)} \,{:=}\,(\mathbf{r}_{1}^{(m)} \,{-}\, \mathbf{r}_{2}^{(m)}) / |\mathbf{r}_{1}^{(m)} \,{-}\, \mathbf{r}_{2}^{(m)}|$.
If the collision happens at the head of the $m^\text{th}$ filament, $\theta_0^{(m)}$ is given by the orientation of its director $ \mathbf{u}_0^{(m)}$. 
The distance vector (red arrow) $\Delta \mathbf{r}_{nm} = (\mathbf{r}_0^{(n)} \,{-}\, \mathbf{r}^{(m)})_\text{shDist}$ is the normal vector to the center-line of filament $m$ between $\mathbf{r}_1^{(m)}$ and $\mathbf{r}_2^{(m)}$, connecting to $\mathbf{r}_0^{(n)}$.}
\label{Fig::A1}
\end{figure}

We consider a system of $M$ polymer filaments, each with a fixed length $L$ and a width $d$.
Individual polymers are modelled as discrete, slender chains consisting of $N\,{-}\,1$ identical cylindrical segments connected by $N$ identical spherical joints; for an illustration see Fig.~\ref{Fig::A1}. 
In this way, each point along the polymer's contour has a well-defined, smooth surface and tangential direction, reducing artificial friction effects due to the discretization present in bead-spring-like representations~\cite{abkenar_collective_2013}.

The polymers perform a trailing motion on a planar surface: as the head of the polymer changes its direction the tail strictly follows the trajectory traced out by the head.
This resembles the typical situation observed in actomyosin motility assays where in a planar geometry actin filaments are propelled along their contour by immobilized molecular motors and where motion orthogonal to the filament contour is suppressed~\cite{schaller_polar_2010, suzuki_polar_2015}.
In these experimental setups, it is observed that the head of each polymer performs a persistent random walk (with persistence length $L_p$), and, in addition, changes its direction due to local alignment interactions when colliding with other polymers. 

In order to model this dynamics, we describe each polymer $n$ by the positions $\mathbf{r}_j^{(n)}$ of its spherical joints $j$, where $n \,{\in}\, \{ 0, 1, \ldots, M \,{-}\, 1 \}$ and $j \,{\in}\, \{ 0, 1, \ldots, N \,{-}\, 1 \}$ (with the head of a polymer denoted by $j\,{=}\,0$); for an illustration see Fig.~\ref{Fig::A1}.
We assume that---given the direction $\mathbf{u}_0^{(n)}$ of a polymer's head---its equation of motion reads: 
\begin{align} 
\label{seq:evo01}
	\partial_t \mathbf{r}_0^{(n)}
	&= v \, \mathbf{u}_0^{(n)} -\mathbf{F_\text{rep}}
	 = v
	 \left(
	 \begin{array}{c}
	 \cos \theta_0^{(n)} \\ \sin \theta_0^{(n)} 
	 \end{array}
	 \right)-\mathbf{F_\text{rep}} 
	 \, . 
\end{align}
Here $\theta_0^{(n)}$ denotes the $n^\text{th}$ polymer's orientation and $v$ the velocity of a free polymer. 
$\mathbf{F_\text{rep}}$ is a weak repulsive force (the exact definition of which we will give later in Eq.~\eqref{seq:repFrc}) which only acts when the filament head overlaps with the head or tail of another polymer. 
The speed $v^{(n)}$ of filament $n$ is given by the absolute value of $\partial_t \mathbf{r}_0^{(n)}$.

The equation of motion for the orientation $\theta_0^{(n)}$ of the $n^\text{th}$ polymer's head is given by 
\begin{align} \label{seq:evo1}
	\partial_t \theta_0^{(n)}
	&=- \frac{\delta H_0^{(n)}}{\delta \theta_0^{(n)}}
	  + \sqrt{\frac{2v }{L_p}} \, \xi \, ,
\end{align}
where the first term denotes the effect of other filaments on the orientation of filament $n$, and $\xi$ is an angular random white noise with zero mean and unit variance; the amplitude of the noise ensures that the value of the path persistence length of a free polymer is given by $L_p$.  
The effective potential $H_0^{(n)}$ acting on the director of filament $n$, is given by a sum $H_0^{(n)}\,{=}\,\sum_{m} U_{m}^{(n)}$ over the alignment potentials $U_{m}^{(n)}$. These potentials describe the alignment interaction between filament $m$ and the head of filament $n$, and will depend on both the relative distance and the relative orientation of these filaments.
To define these potentials we introduce the distance vector [Fig.~\ref{Fig::A1}]
\begin{equation}
	\Delta \mathbf{r}_{nm}
	= 
	\Bigl(
	\mathbf{r}_0^{(n)} - 
	\mathbf{r}^{(m)}
	\Bigr)_\text{shDist}
	\, ,
\end{equation}
which denotes the vector connecting the head of polymer $n$ with that part of the body (contour) of an adjacent polymer $m$ that has the shortest possible distance to the head [red arrow in Fig.~\ref{Fig::A1}].
We signify the segment $j$ on filament $m$ that filament $n$ collides with as \textit{collision segment}.
The corresponding orientation of this collision segment is denoted by $\theta_j^{(m)}$ [Fig.~\ref{Fig::A1}]. With these definitions, we can now define the alignment potential as
\begin{align}  \label{seq:align1}
	U_{m}^{(n)}
	&= 
	C\left(
	\left|\Delta \mathbf{r}_{nm}\right|
	\right) 
	\times
	\Bigl(
	A_p 
	\left(\Delta \theta_{nm} \right)
	+
	A_n
	\left(\Delta \theta_{nm} \right)
	\Bigr)
	\, ,
\end{align}
where $\Delta \theta_{nm} = \theta_0^{(n)}{-}\theta_j^{(m)}$ denotes the impact angle of the collision of the head of polymer $n$ with the body of filament $m$.
The first factor $C\left(
	\left|\Delta \mathbf{r}_{nm}\right|
	\right)$ accounts for the spatial dependence of the potential.
For simplicity, we assume a potential that vanishes outside of an interaction radius $d$ and increases linearly for smaller distances:
\begin{align} \label{seq:ramp}
	C\left(
	\left|\Delta \mathbf{r}_{nm}\right|
	\right) 
	=
	\left
	\lbrace
	\begin{array}{cc} 
		0 & \text{if}\ 	
		\left|\Delta \mathbf{r}_{nm}\right|{>}d \\ 
		(d-	\left|\Delta \mathbf{r}_{nm}\right|)/d & \text{else} 
	\end{array} \right. 
	\,.
\end{align}
The second factor is a sum of functions $A_{p/n}$ that describe the polar/nematic alignment-torques present during a collision. They are given by
\begin{subequations}
\label{seq:alignTorq}
\begin{align} 
	A_p(\phi) \label{seq:alignTorqP}
	&=
	-\frac{\varphi_p v^{(n)}}{d}\cos \phi 
	\, , \\
	A_n(\phi)  \label{seq:alignTorqN}
	&=
	-\frac{\varphi_n v^{(n)}}{d}\cos 2\phi
	\, ,
\end{align}
\end{subequations} 
with the amplitudes $\varphi_{p/n}$ characterizing the typical angular displacement in a single collision (see Supplemental Material of Ref.~\cite{huber_emergence_2018}).
A variation of $\varphi_{p/n}$ allows to independently and continuously vary the preferences for polar or nematic alignment.
As was shown in Ref.~\cite{huber_emergence_2018}, the WASP simulation model shows the formation of both polar and nematic patterns, depending primarily on the relative alignment strength $\alpha\,{=}\,\varphi_n/\varphi_p$.

To prevent an unphysical aggregation of filaments---that can be triggered by the alignment torques when too many filaments overlap at the same location---we added a very weak repulsion force $\mathbf{F_\text{rep}}$ to Eq.~\eqref{seq:evo01}. It is given by
\begin{align} 
\label{seq:repFrc}
\mathbf{F_\text{rep}}
	&= 
	-s \sum_{m} 
	C
	\left(
	\left|\Delta \mathbf{r}_{nm}\right|
	\right)	
	 \frac{\Delta \mathbf{r}_{nm}}{\left|\Delta \mathbf{r}_{nm}\right|}
	\, ,
\end{align}
where $s \,{\ll}\, 1$ denotes the small amplitude.

In actomyosin motility assays~\cite{schaller_polar_2010, butt_myosin_2010, hussain_spatiotemporal_2013, suzuki_polar_2015,huber_emergence_2018} one observes that the polymer tails follow the movement of their respective filament heads.
In our agent-based model, we emulate this trailing motion as follows:
First, in order to assure tangential motion, for a given filament $n$, each joint $\mathbf{r}_j^{(n)}$ in its tail ($j \,{>}\, 0$) is assumed to move in the direction of 
	$\frac12\, 
	(
	\mathbf{u}_{j+1}^{(n)} \,{+}\, \mathbf{u}_j^{(n)}
	)$, corresponding to the average of the segment's orientations adjacent to that joint [see Fig.~\ref{Fig::A1}].
Second, to also maintain an average length $b$ of the cylindrical segments between the bonds
we assume a linear (Hookian) restoring force with spring coefficient $K_s$. 
Taken together, the equation of motion of a tail joint $j$ is defined as
\begin{align}
	\partial_t \mathbf{r}_j^{(n)}
	=
	 K_s \, 
	\left(
	\left| \mathbf{r}_{j}^{(n)}{-}\mathbf{r}_{j-1}^{(n)}\right| - b
	\right) \,
	\frac12
	\left(
	\mathbf{u}_{j+1}^{(n)}+\mathbf{u}_j^{(n)}
	\right)
	\, .
\end{align}
We chose $K_s \,{=}\, 200$  sufficiently large to keep the cylinder length close to its average value  $b$.

In our simulations we observed that the performance of our algorithm significantly depended on the number of times the alignment torques,  Eq.~\eqref{seq:alignTorq}, were calculated. 
We, therefore, were searching for an averaging scheme that would reduce the computation of the alignment torques to at most once per filament per time step.
The main idea put forward in Ref.~\cite{huber_emergence_2018}---and also shown there not to affect the system's dynamics---is to implement an averaging scheme as follows:
One replaces the sum in $H_0^{(n)}$ by an averaged quantity $\tilde{H}_0^{(n)} $ defined as
\begin{align} \label{seq:align22}
	\tilde{H}_0^{(n)}
	&=  
	A_p
	\bigl(
	\Delta \theta^{(n)}_{p} 
	\bigr) 
	|\text{q}_p|
	+
	\tilde{A}_n
	\bigl(
	\Delta \tilde{\theta}^{(n)}_{n}
	\bigr) 
	|\Delta\widetilde{\mathbf{e}}_{n}|
	\, .  
\end{align}
The first term in Eq.~\eqref{seq:align22} (polar interaction)  is motivated as follows:
Instead of calculating the polar torques,  Eq.~\eqref{seq:alignTorqP}, for each adjacent polymer $m$ and then summing over all these polymers with weights given by the repulsive linear potential $C (|\Delta \mathbf{r}_{nm}|)$, we determine the quantity 
\begin{align} \label{seq:q2}
	\text{q}_p 
	= 
	\sum_{m}  
	C\left(
	\left|\Delta \mathbf{r}_{nm}\right|
	\right)  
	\frac{v^{(m)}}{v}  e^{i  \theta_j^{(m)} } 
	\, .
\end{align} 
It defines the average in the velocities of all the collision segments $j$ over all filaments $m$ weighted by the strength of the impact, $C\left(
	\left|\Delta \mathbf{r}_{nm}\right|
	\right)$, of filament $n$ with them.
In other words, this vector characterizes the weighted (by interaction strength) average of the velocities of the collision segments.
We then use the orientation $\theta_{p} \,{=}\, \text{arg}(\text{q}_p)$ of the average velocity to calculate the average excerted torque, $A_p \bigl(
	\Delta \theta^{(n)}_{p} 
	\bigr) $ using the average polar impact angle defined as $\Delta \theta^{(n)}_{p} \,{=}\,  \theta_0^{(n)} \,{-}\, \theta_{p}$. 
Note that the magnitude of $\text{q}_p$ measures the average strength of all the polar impacts on filament $n$.
Here we have additionally introduced a velocity dependence ($v^{(m)}/v$ in  Eq.~\eqref{seq:q2}) to emulate that polar alignment in the motility assay is mainly caused by friction between filaments. 
With this, our agent based model can also be used in cases where filament velocities are broadly distributed. 
Since the filament velocity in the present study is constant and only very weakly influenced by $\mathbf{F_\text{rep}}$, this velocity dependence can also be omitted without affecting the results.

The second term in Eq.~\eqref{seq:align22} is motivated in a similar fashion as the first one:
Instead of calculating  Eq.~\eqref{seq:alignTorqN} for each adjacent polymer $m$, we define a weighted average direction of the connecting vector $\Delta\widetilde{\mathbf{e}}_{n}$ 
\begin{align} \label{seq:q1}
	\Delta\widetilde{\mathbf{e}}_{n}
	:=  
	\sum_{m} 
	C\left(
	\left|\Delta \mathbf{r}_{nm}\right|
	\right) 
	\frac{\Delta \mathbf{r}_{nm}}
		 {\left|\Delta \mathbf{r}_{nm}\right|}
	\, .
\end{align} 
weighted, again, by the strength of the respective impact.

The overall magnitude of the repulsive potential to nematic alignment is given by the absolute value of $\Delta\widetilde{\mathbf{e}}_{n}$. 

Similarly as for the polar case, we used the orientation $\tilde{\theta}_{n}$ of the vector $\Delta\widetilde{\mathbf{e}}_{n}$ to define an average nematic impact angle as $\Delta \tilde{\theta}^{(n)}_{n} \,{=}\,  \theta_0^{(n)} \,{-}\, \tilde{\theta}_{n}$, which we used to compute the average nematic alignment torque in Eq.~\eqref{seq:align22}. 
Note that the nematic term in Eq.~\eqref{seq:align22} reads
\begin{subequations}
\begin{align}
	\tilde{A}_n(\theta)
	&= \frac{\varphi_n v^{(n)}}{d} \cos 2\theta,
\end{align} 
\end{subequations}
 since $\tilde{\theta}_{n}$ is derived from the normal vectors to the polymer contours (and not the tangential vectors, as it was done before).

\subsection{WASP implementation and parameters} \label{sup:wasp_implementation_and_parameters} 
Algorithmically, we integrate the dynamics by a straightforward Euler algorithm, which was implemented in C++ using a heavily parallelized architecture in OpenMP~\cite{dagum1998openmp}. 
Maximal performance of the simulation was achieved by employing a cell algorithm and Verlet lists~\cite{wang_algorithm_2007} that exploit the fact that filament interactions are short-ranged. 
This implementation resulted in a practically linear scaling of simulation times with $M$ (the number of filaments in the system).
Throughout this work and if not stated otherwise, we fixed some of the model parameters to values similar to those used in Ref.~\cite{huber_emergence_2018}: filament aspect ratio $L/d\,{=}\,21$, discretization $N\,{=}\,5$, persistence length $L_p\,{=}\,31.75L$, and velocity $v\,{=}\,1$. 
The polar alignment strength was fixed to $\varphi_p\,{=}\,0.036\approx2.1^\circ$ to obtain collision statistics similar to those observed experimentally~\cite{suzuki_polar_2015}. 
Moreover, we used a system consisting of $10^4$ filaments and a periodic simulation box of length $L_\text{box}\,{=}\,81.3L$. 
Simulations were started with random initial conditions, i.e.\ filaments were placed at random positions and with random orientations in the simulation box.
Time is measured in units of the correlation time $L_p/v$.

\subsection{Cluster polar order and other order parameters}\label{sup:cluster_polar_order_and_other}  
As described in the main text, we decomposed the assembly of polymers into clusters of close-by polymers. 
To that end, we define the distance between two polymers $n$ and $m$ as the length of the shortest one of the set of distance vectors $\mathbf{r}_j^{(n)}-\mathbf{r}_i^{(m)}$ between their nodes $j$ and $i$.
We calculated all distances between adjacent polymers, and assigned polymers to the same cluster if their distance was smaller than the bond length $b$. 

Next, to properly define the degree of polar order for each of these clusters, we defined the net polar order of a cluster (of size $k$) as $\pi_k \,{:=}\, p_k \,{-}\, \Delta_k$, where $\Delta_k$ denotes the expected nonzero polar order of clusters where the orientation of each filament is chosen at random; the cluster polar order was defined as $p_k \,{:=}\, \frac{1}{k}|\sum_{j=1}^k \exp(i\theta_j)|$.
The quantity $\Delta_k$ is obtained by calculating the mean polar order $\Delta_k=\frac{1}{k}\langle |\sum_{j=1}^ke^{iO_j}| \rangle $ with the filaments' orientations $O_j$ uniformly distributed in the interval $[-\pi;\pi]$.
Explicitly writing out the absolute value, $\Delta_k$ reads
\begin{align}
	\Delta_k
	=
	\frac{1}{k}
	\Bigl\langle 
	 \Bigl| \sum_{j=1}^ke^{iO_j} \Bigr| 
	\Bigr\rangle  
	=
	\frac{1}{k}
	\Bigl\langle
	\Bigl( \! \sum_{m,n=1}^k  
	       e^{i(O_m-O_n)} 
	\Bigr)^{1/2}
	\Bigr\rangle 
	\, . 
\end{align}
By splitting up the double sums and introducing the shorthand notation $\delta_m^n\,{=}\,O_m \,{-}\, O_n$, this can be further rewritten as
\begin{align}
	\Delta_k
	=
	\frac{1}{k}
	\Bigl\langle
	\Bigl( 
	\sum_{m=n}^k 1 
	+    
	\sum_{m=1}^k \sum_{n=m+1}^k e^{i\delta_m^n}  
	+   
	\sum_{m=1}^k \sum_{n=1}^{m-1} e^{i\delta_m^n}    
	\Bigr)^{1/2}
	\Bigl\rangle
	\, .
\end{align}
Evaluating the first sum and renaming the indices in the last sum, one obtains
\begin{align}
	\Delta_k
	=
	\frac{1}{k}
	\Bigl\langle
	\Bigl(  
	k 
	+    
	\sum_{m=1}^k \sum_{n=m+1}^k e^{i\delta_m^n}  
	+   
	\sum_{m=1}^k \sum_{n=m+1}^{k} e^{-i\delta_m^n}    
	\Bigr)^{1/2}
	\Bigl\rangle
	\, .
\end{align}
With the shorthand notation $\sum_{m=1}^k\sum_{n=m+1}^k \,{=:}\, \sum_{(m,n)}$ this can be written as 
\begin{align}
	\Delta_k
	=
	\frac{1}{k}
	\Bigl\langle
	\Bigl(  
	k 
	+ 
	2\sum_{(m,n)} \cos \delta_m^n
	\Bigr)^{1/2}
	\Bigl\rangle
	\, .  
\end{align}
Finally, by expanding the square root in powers of $\cos \delta_m^n$ one finds
\begin{align}\nonumber
	\Delta_k
	=& 
	\frac{1}{k} 
	\Bigl\langle 
	\sqrt{k}
	+
	\frac{1}{\sqrt{k}} \sum_{(m,n)} \cos \delta_m^n 
	- 
	\frac{1}{2k^{3/2}}\sum_{(m,n)} \cos^2 \delta_m^n 
	\\ 
	\label{seq:randval}
	& 
	+ 
	\frac{1}{2k^{5/2}}\sum_{(m,n)} \cos^3 \delta_m^n 
	+ \mathcal{O}(k^{-3/2})
	\Bigr\rangle
	\, .
\end{align}
Since $\langle \cos^j \delta_m^n\rangle\,{=}\,0$ for $j$ odd and $\langle \cos^2\delta_m^n\rangle\,{=}\,\frac{1}{2}$, this can be further simplified (note that, for $n\,{\ne}\,u$ or $m\,{\ne}\,v$, terms of the form $\langle\sum_{(u,v)}\sum_{(m,n)} \cos \delta_u^v \cos \delta_m^n \rangle$ can be factorized and thereby give no contribution in Eq.~\eqref{seq:randval}).
By evaluating the remaining sum, one obtains 
\begin{align} \nonumber
	\Delta_k
	&= 
	\frac{1}{\sqrt{k}}\left(1 - \frac{(k-1)}{8k}\right)+\mathcal{O}(k^{-5/2}) \\ \label{seq:randvalshort}
	&= \frac{1}{\sqrt{k}}\left(\frac{7}{8}+\frac{1}{8k}\right) +\mathcal{O}(k^{-5/2}).
\end{align}

\begin{figure}[b]
\centering
\includegraphics[width=1.\columnwidth]{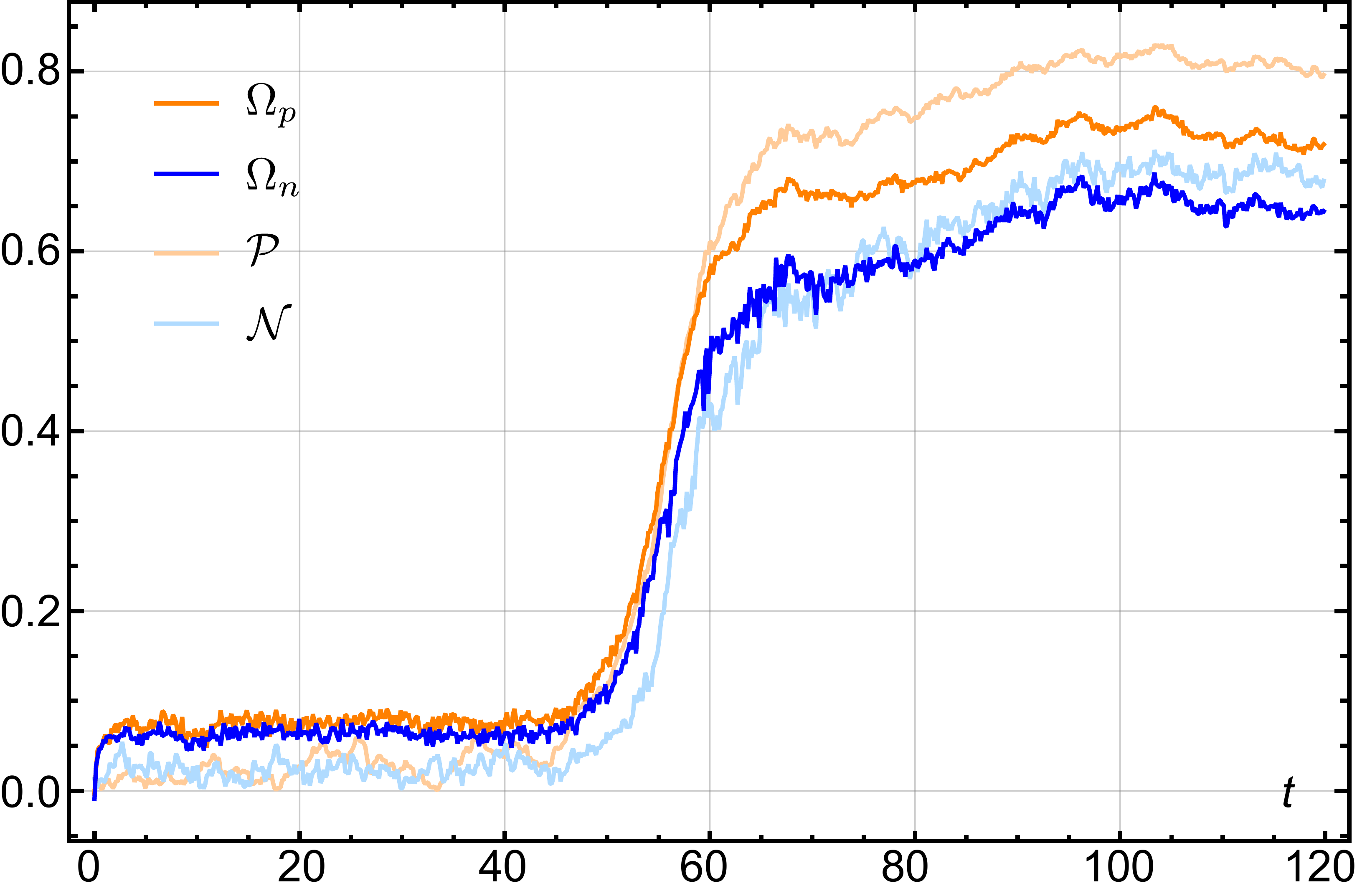}
\caption{
\textbf{Global order parameters:} 
Temporal evolution of the global \textit{polar} order  parameters $\Omega_p$ and $\mathcal{P}$, and the global \textit{nematic} order parameters $\Omega_n$ and $\mathcal{N}$, as indicated in the graph. Parameters: $\alpha\,{=}\,2$.}
\label{Fig::A2}
\end{figure}

In the main text, we defined the \textit{cluster polar order parameter} as an average of the net order $\pi_k$ weighted by the respective cluster size $k$:
\begin{equation}
	\Omega_p \,{:=}\,  \tfrac{1}{M} \sum_{\{c\}}\pi_k^{(c)} k^{(c)}
\end{equation}

This has to be distinguished from the alternative definition of a global polar order parameter 
\begin{equation}
	\mathcal{P} 
	=
	\frac{1}{M} 
	\Bigg|
	\sum_{j=0}^{M-1}e^{i\theta_0^{(j)}}
	\Bigg| 	
	\, ,
\end{equation}
which is an average over all filament orientations independent of which clusters they belong to.
The temporal evolution of both of these global order parameters, $\Omega_p$ and $\mathcal{P}$, is shown in Fig.~\ref{Fig::A2}. 
Although they are related quantities, there are clear differences: 

(i) In the disordered phase, $\Omega_p$ still displays a nonzero value stemming from the small average polar order of the clusters present in the system. In contrast, $\mathcal{P}$ is almost zero in the disordered phase as it is averaged over all filaments in the system, whose orientations cancel out. 
(ii) In the ordered phase, however, $\Omega_p$ is smaller than $\mathcal{P}$ as single `ordered' clusters are not contained in the sum for $\Omega_p$; note that $\pi(1)\,{=}\,0$.
Throughout this work we prefer to use $\Omega_p$, since it is more sensitive to polar structures which form in independent parts of a system, but whose orientations are not yet correlated. 
For example, two non-overlapping polar clusters of the same size and order, but opposite orientations, would yield $\mathcal{P}\,{=}\,0$, whereas their presence would be detected with $\Omega_p$.  

Similarly, one can define two distinct types of nematic order parameters, $\Omega_n$ and $\mathcal{N}$, by simply replacing every angle $\theta$ with $2\theta$ in the above definitions; see Fig.~\ref{Fig::A2} for an example. 
However, since in our study we only investigate polar structures and in this case the nematic order parameter is slaved to the polar order, it is of little importance for our analysis.

\subsection{Time scale analysis}\label{sup:time_scale_analysis}
\subsubsection{Measurement of $t_0$, $t_d$ and $\tau $}
To obtain the initial time scale $t_0$, the dwell time $t_d$, and growth time $\tau$ from our data, we analysed the temporal evolution of the cluster polar order parameter $\Omega_p (t)$ (Fig.~\ref{Fig::A2}). 
To this end, we looked for a fit function $f(t)$ for $\Omega_p (t)$, which should capture  the main features of its temporal evolution: 
(i) fast rise towards the quasi-stationary, disordered regime (within a short time $t_0$),  
(ii) plateau until $t_d$, 
(iii) exponential growth starting at time $t_d$. 
In our analysis we decided to use the following piecewise defined function
\begin{align}
\label{seq:fitFktn} 
	f(t)
	=
	\left\lbrace
	\begin{array}{cc}
	a \left(1- e^{-t/t_0} \right)  
	&\text{for} \;\, t<t_d \\
	a \left(1- e^{-t_d/t_0}\right) e^{(t-t_d)/\tau}  
	&\text{for} \;\,  t>t_d .
\end{array}\right. 
\end{align}
Here $a$ is a fit parameter that quantifies the small, yet nonzero value of $\Omega_p$ during the quasi-stationary, disordered regime before nucleation. 
The fit was made up to the time point at which $\Omega_p(t)>0.5$ for the first time, that is before $\Omega_p(t)$ started to saturate again.

\subsubsection{System size dependence of $t_d$ and $\tau $}
\begin{figure}
\centering
\includegraphics[width=1.\columnwidth]{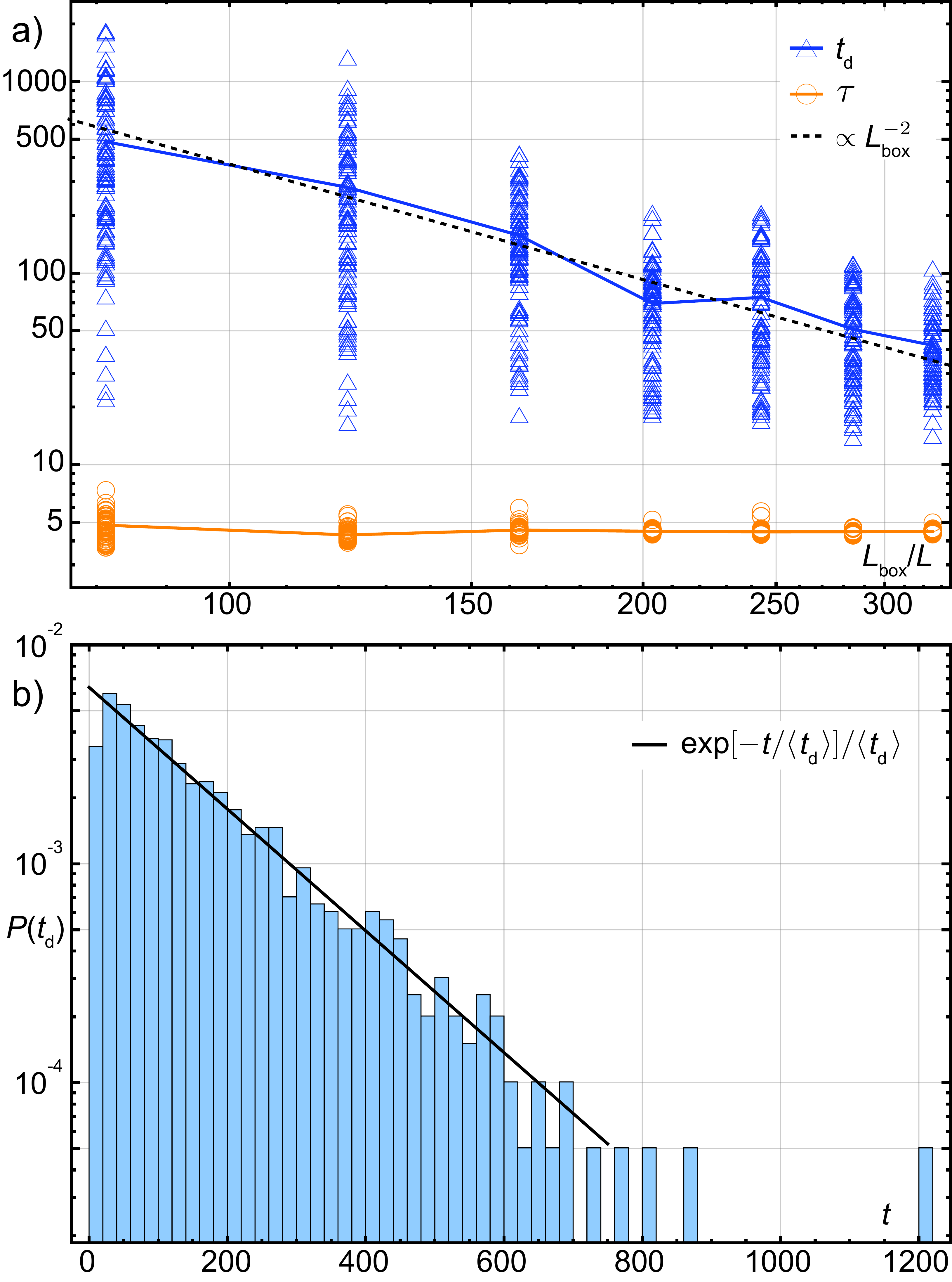}
\caption{
\textbf{System size dependence and distribution of waiting times.} 
(a) Waiting time $t_d$ and growth time $\tau$ as a function of the system size $L_\text{box}$ in units of the polymer length $L$. Solid lines denote average values (taken over $90-100$ independent simulations for each system size); data are shown as triangles and circles for $t_d$ and $\tau$, respectively. The black dashed line indicates a scaling law proportional to the area of the system. 
(b) Histogram of waiting times $t_d$ taken over an ensemble of 1000 simulations. The black solid line shows an exponential waiting time distribution $P(t_d)$ with mean $\langle t_d \rangle \,{=}\,156$. Parameters: $\rho L^2\,{=}\,1.51$, $\alpha\,{=}\,1.583$ for (a) and $\alpha\,{=}\,1.67$, $L_\text{box}\,{=}\,81.3L$ for (b).}
\label{Fig::A3}
\end{figure}

In the main text, we studied how the characteristic times $t_d$ and $\tau $ depend on the relative alignment strength $\alpha $ [Fig.~\ref{Fig::2}(d)].
Here, we additionally investigate how these quantities depend on the system size; see Fig.~\ref{Fig::A3}(a). 
We find that the expected dwell time $\langle t_d \rangle$ scales inversely with the area of the system, $L_\text{box}^2$. 
This indicates that---for each given set of parameters---there is a constant probability per unit of area to nucleate a cluster large enough to trigger the exponential increase of order in the system. 
Hence, the formation of critical nuclei occurs independently in different parts of the system.

We further observe that the growth time $\tau $ is approximately independent of system size ($\tau\approx 4.5$), although $L_\text{box}$ is increased by more than a factor of $3$. 
This is probably caused by the fact that on the one hand the mass of ordered clusters growths (after a critical nucleus has formed) exponentially with time, but that on the other hand the total filament mass of ordered clusters in an ordered system  (i.e. the mass that has to be incorporated into the ordered clusters during the growth process) grows only approximately proportionally to the size of the system. 
It therefore takes only a very short time for the additional filaments (introduced by the increase in system size) to be incorporated into the ordered clusters.
Hence, in order to observe a significant change of $\tau $, one would have to increase the number of filaments in the system (and thus $L_\text{box}$) by far more than a small factor; this however is beyond the numerically feasible limit.

\subsubsection{Variance of the nucleation time}\label{sup:variance_of_the_nucleation_time}
We also recorded the statistics of nucleation times $p(t_d)$ at one point in parameter space and for a small value of $\alpha$ (Fig.~\ref{Fig::A3}(b)).
Similar as in classical nucleation theory~\cite{sear_nucleation_2007, kalikmanov_nucleation_2013}, it exhibits an exponential distribution of times.
\begin{figure}
\centering
\includegraphics[width=1.\columnwidth]{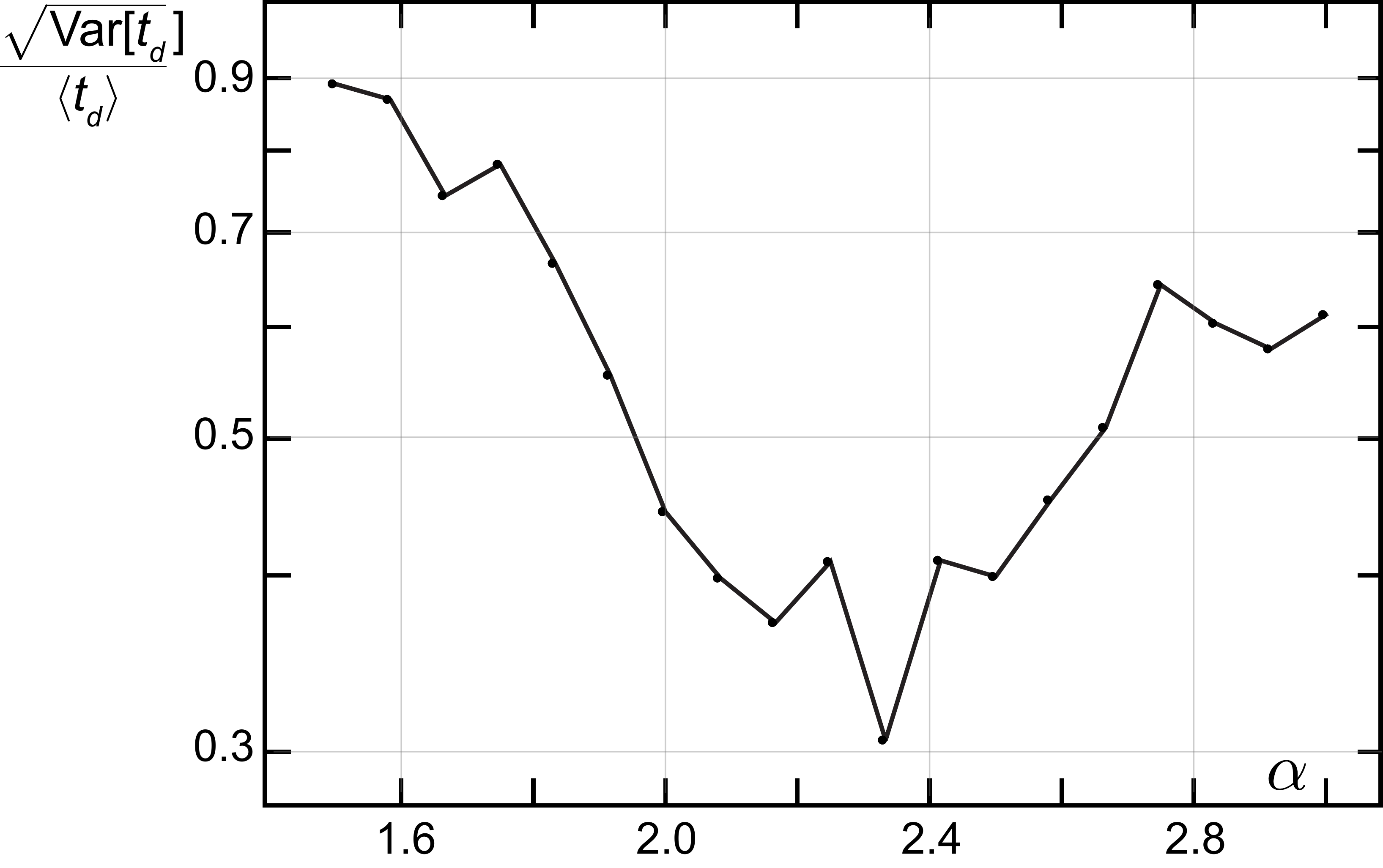}
\caption{
\textbf{Coefficient of variation} for the dwell time $t_d$, $CV \,{=}\, \sqrt{\mathrm{Var}[t_d]}/\langle t_d\rangle$, as a function of $\alpha$. Parameters and data are identical to Fig.~\ref{Fig::2}(d).}
\label{Fig::A4}
\end{figure}
This is also reflected in the coefficient of variation $CV \,{=}\, \sqrt{\mathrm{Var}[t_d]}/\langle t_d\rangle \,{\approx}\, 1$, see Fig.~\ref{Fig::A4}.
With increasing $\alpha$, however, we observe that the average dwell time $t_d$ shrinks until eventually the system instantly starts to develop polar order  [cf. Fig.~\ref{Fig::2}(d)].

This decrease of $t_d$ is accompanied by a decrease of the coefficient of variation [Fig.~\ref{Fig::A4}], indicating that the waiting times are no longer exponentially distributed. 
Since  $t_d$ would always be zero in the limit of an instantaneous nucleation (and would also not fluctuate any more), this is in accordance with the above observation.

The subsequent increase of the coefficient of variation (after the minimum at $\alpha \,{\approx}\, 2.3$) is an artefact. 
It can be attributed to an increased error of the fit used to determine $t_d$; cf.  Eq.~\eqref{seq:fitFktn}. 
This increase in error is due to the fact that $\Omega_p$ no longer shows a clear plateau after reaching the metastable state but instead directly continues to grow exponentially towards macroscopic order.  

\subsection{Cluster stability analysis}\label{sup:cluster_stability_analysis}
\subsubsection{Critical polar moment}\label{sup:critical_polar_moment}
As discussed in the main text, we probed the stability of the disordered state by inserting perfectly ordered clusters of size $\tilde{k}$ (and hence polar moment $S\,{=}\,\tilde{k}$) into systems at time points where they were still in the metastable disordered state. 
Specifically, we chose the time point $t\,{=}\,5 > t_0$, sufficiently later than the time when the systems had reached the metastable state. 
To keep the overall filament density constant, we extracted $\tilde{k}$ filaments at random and used them to construct the clusters with which we probed the system. 
To this end, we stacked these filaments in parallel, with a transversal distance $d$; see Fig.~\ref{Fig::3}(a) for an illustration of such a cluster. 
We inserted the so formed cluster at a randomly chosen  position and with random initial orientation.

\begin{figure}
\centering
\includegraphics[width=1.\columnwidth]{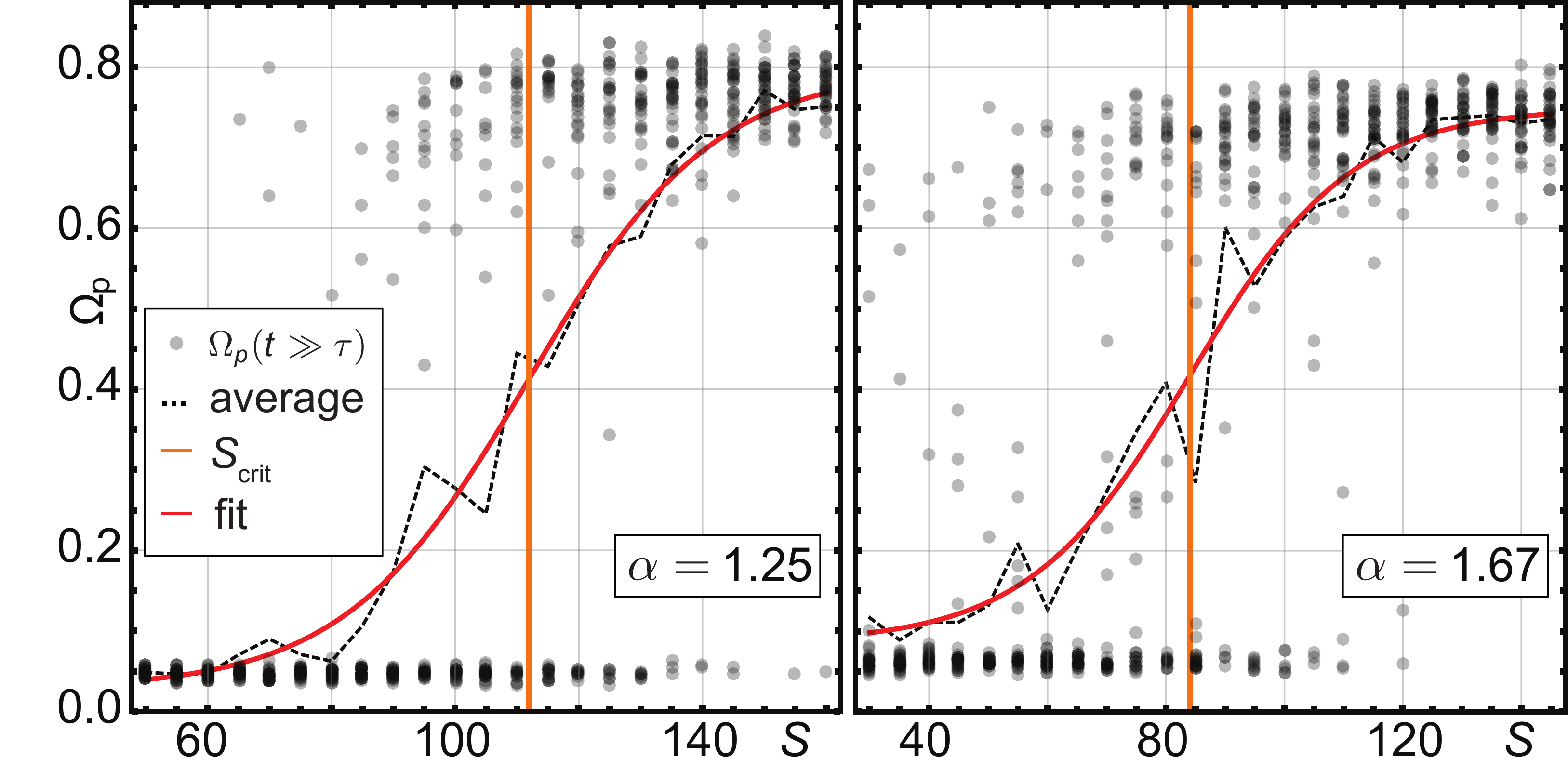}
\caption{
\textbf{Cluster stability analysis.} 
Scatter plots of the cluster polar order parameter $\Omega_p$ at $t \,{=}\, 25 \,{\gg}\, \tau$ (gray open circles) for different initial sizes $S$ of artificially inserted perfectly ordered clusters; for each set of parameters we conducted $30$ independent simulations runs.
The left and right panel show the results for relative alignment strengths $\alpha \,{=}\, 1.25$ and for $\alpha \,{=}\, 1.67$, respectively.
The dashed black and solid red line indicate the average of the polar order parameter and a sigmodial fit, respectively. 
Vertical orange lines indicate the approximate values for the critical polar moment $S_\text{crit}$.}
\label{Fig::A5}
\end{figure}

We then monitored the temporal evolution of the cluster polar order parameter $\Omega_p$ until a given time point $t\,{=}\,25$, which we chose such that it is much larger than $\tau$. 
Figure~\ref{Fig::A5} shows a scatter plot of 30 realizations for each set of parameters as a function of $S$, for two different values of the relative alignment strength $\alpha$. 
As can be inferred from the statistical distribution of the observed cluster polar order parameters $\Omega_p$ (at times $t \gg \tau)$, there is no hard threshold for the cluster size above which the system always develops polar order.
Instead, the probability that insertion of the artificial nucleation seed leads to order formation increases gradually with $S$ over some finite width. 
We define the critical value $S_\text{crit}$ as that value of $S$ which leads to the emergence of polar order with probability $\frac12$. 
To determine $S_\text{crit}$ from the recorded data, we fitted the averaged polar order parameter (which is proportional to the nucleation probability) with a sigmoid function of the form 
\begin{align}
	f(S) 
	= 
	a + 
	\frac{b}{1+e^{-(S-S_\text{crit})/c}}
	\, ,
\end{align}
where $a$, $b$ and $c$ are fitting parameters (Fig.~\ref{Fig::A5}). 
 \begin{figure}
\centering
\includegraphics[width=1.\columnwidth]{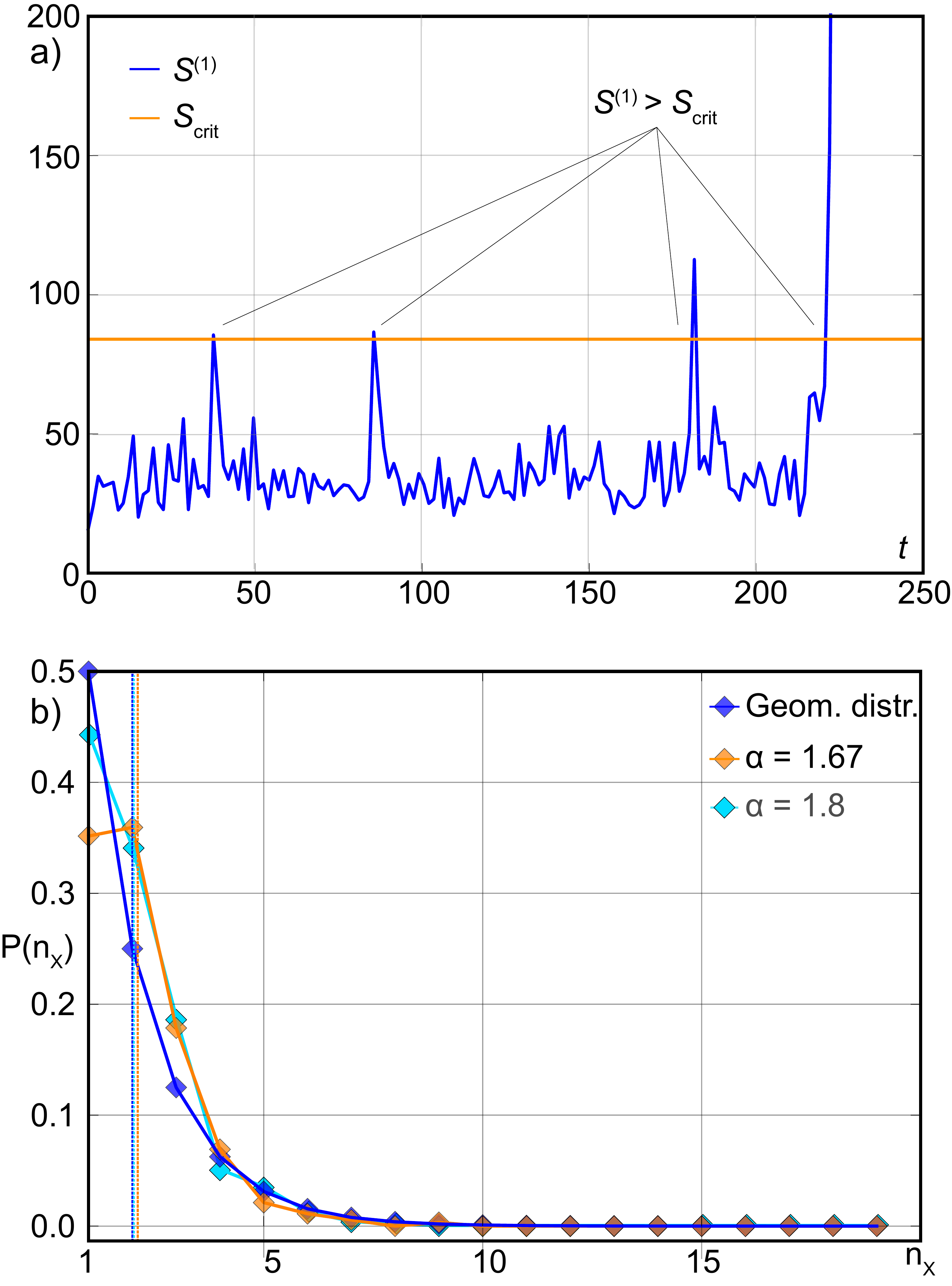}
\caption{
\textbf{Spontaneous formation of critical nucleation clusters.}
\textbf{(a)} 
Temporal evolution of the largest polar moment $S^{(1)}$ in a single simulation run (blue solid line). 
The orange line represents the value of the critical polar moment $S_\text{crit} \,{\approx}\, 84$ as obtained from Fig.~\ref{Fig::A5} (right panel).
Before the system eventually exhibits rapid formation of polar order, there are three instances where it crosses that line but is not successful in developing polar order. 
Parameters: $\alpha \,{=}\, 1.67$, $\Delta t \,{=}\, 1.5$.
\textbf{(b)} 
Probability distribution $P(n_\times)$ of the number of times $n_\times$ the  largest polar moment $S^{(1)}$ exceeds the threshold $S_\text{crit} $ before it finally succeeds in forming polar order, obtained from the simulation data (orange and cyan line, for $\alpha \,{=}\, 1.67$ and $\alpha \,{=}\, 1.8$, respectively), in comparison with a geometric distribution (blue line) with parameter $p \,{=}\, 0.5$. 
The dashed vertical orange and cyan line represent the mean value of the simulation data ($\langle n_\times  \rangle \,{=}\, 2.125 $ and  $ \langle n_\times  \rangle \,{=}\, 2.02 $), for $\alpha \,{=}\, 1.67$ and $\alpha \,{=}\, 1.8$, respectively.
The expectation value of the geometric distribution for $p \,{=}\, 0.5$ ($\mathbb{E}(n_\times) \,{=}\, 2$) is shown as a blue vertical line. 
Data were obtained in $892$ simulation runs for each $\alpha$. For $\alpha \,{=}\, 1.8$ $S_\text{crit} \,{\approx}\, 75$ was obtained with the same method as shown in Fig.~\ref{Fig::A5} (data not shown).
}
\label{Fig::A6}
\end{figure}

\subsubsection{Critical polar moment and spontaneous nucleation}\label{sup:critical_polar_moment_and_spont}

We have tested whether the value of $S_\text{crit}$---as obtained by insertion of artificial seeds---faithfully predicts the nucleation threshold for the spontaneous formation of polar order.
To this end, we performed simulations in a parameter range where $t_d$ is small; see Fig.~\ref{Fig::A6}(a) for a single simulation run for $\alpha \,{=}\, 1.67$.
As can be inferred from this figure, the cluster with the largest polar moment $S^{(1)}$ needs several `attempts' before it finally succeeds in triggering the formation of polar order in the system.
Given a threshold value $S_\text{crit}$, one expects that each time $S^{(1)}$ exceeds this threshold it leads to polar order formation only with a certain success probability $p_\text{crit}$.
This implies that---sampling over many realization---the number of attempts $n_\times$ needed to trigger formation of polar order is given by a geometric distribution,
\begin{equation}
	P(n_\times) 
	= 
	p_\text{crit} \, (1-p_\text{crit})^{n_\times-1} 
	\, .
\end{equation}
We define the critical cluster size such that if a cluster with a polar moment $S_\text{crit}$ is formed randomly it should---on average---in half of the cases lead to the formation of polar order, i.e.\ the success probability should be $p_\text{crit} = 0.5$.

Indeed, our simulations show that the success probability closely resembles a geometric distribution [Fig.~\ref{Fig::A6}(b)];
for two values of $\alpha$ we sampled over $892$ realizations with different random initial conditions and the same threshold value as found in the simulations using artificially inserted clusters.
Moreover, the geometric distribution and the histogram obtained from our simulation data show the same mean value.

\subsubsection{Course of nucleation in k-p space}\label{sup:course_of_nucleation_in}
\begin{figure}[t]
\centering
\includegraphics[width=1.\columnwidth]{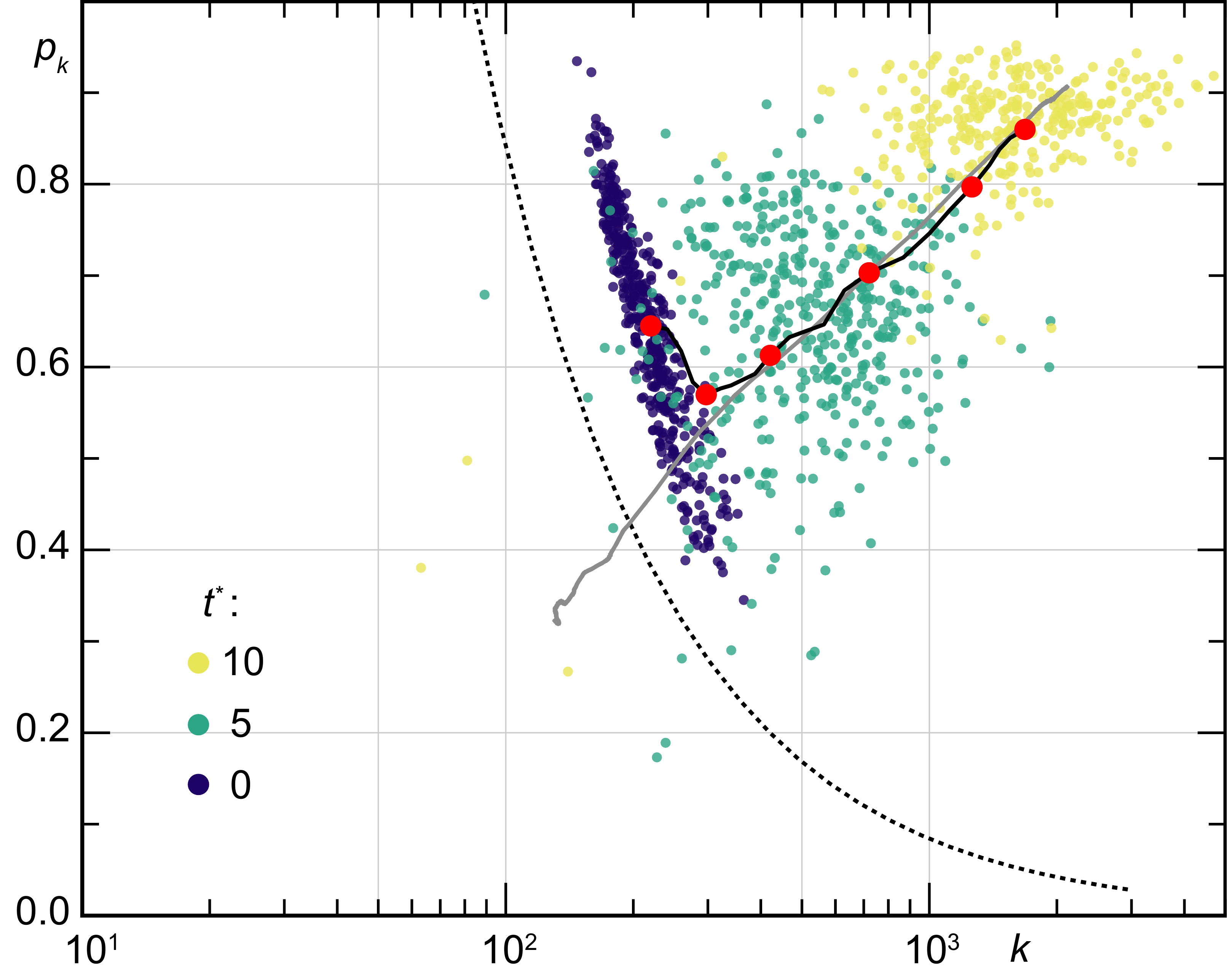}
\caption{
\textbf{Scatter plot for the size $k$ and order $p$ of the clusters with the largest polar moment $S^{(1)}$}.
At time point $t^* \,{=}\, 0$, perfectly ordered clusters of size $k \,{=}\, 140$ are inserted into the system and their temporal evolution is monitored for $431$ independent realizations. 
The ensuing probability clouds at different time points $t^*$ are indicated in the graph with different color. 
As time progresses the cloud moves on average along a trajectory indicated by the black solid line, which depicts the average path of $\langle S^{(1)}\rangle$ in $k$-$p$ space; for comparison the average path from Fig.~\ref{Fig::2}(f) is shown in gray.
The red circles mark the average $\langle S^{(1)}\rangle$ at equidistant timepoints ($\Delta t \,{=}\, 2$, starting at $t^* \,{=}\, 0$). The dashed line indicates $S_\text{crit} \,{\approx}\, 84$.  Same parameters as for Fig.~\ref{Fig::2}(f).}
\label{Fig::A7}
\end{figure}
As discussed in the main text and shown in Fig.~\ref{Fig::2}(f) we monitored and sampled the temporal evolution of clusters with the largest polar moment $S^{(1)}$ in $k{-}p$ space for a sample size of $892$ independent realizations. 
We tested whether our agent-based simulations take the same path towards polar order also if nucleation is triggered by insertion of an artificial nucleation seed, instead of waiting for a spontaneous nucleation and growth event to happen [Fig.~\ref{Fig::2}(f))].
To this end, we inserted perfectly ordered clusters of size $k \,{=}\, 140$ into $431$ different systems at a time point $t\,{=}\,5$ where the system was still in a disordered state.
As can be inferred from Fig.~\ref{Fig::A7}, the probability cloud of $S^{(1)}$ values rapidly becomes indistinguishable from the cloud shown in Fig.~\ref{Fig::2}(f).
Moreover, the center of mass follows, after some initial transient,  the same path as the center of mass of clusters in systems where these clusters spontaneously emerged.
Note that the linear spread of the cloud at $t^* \,{=}\, 0 $ is due to an overlap of the perfectly ordered seeds (placed into the system) with disordered clusters (already present in the system).

\subsection{Steady-state flux of the flocking state}\label{sup:steady_state_flux_of_the_flocking}
\begin{figure}[t]
\centering
\includegraphics[width=1.\columnwidth]{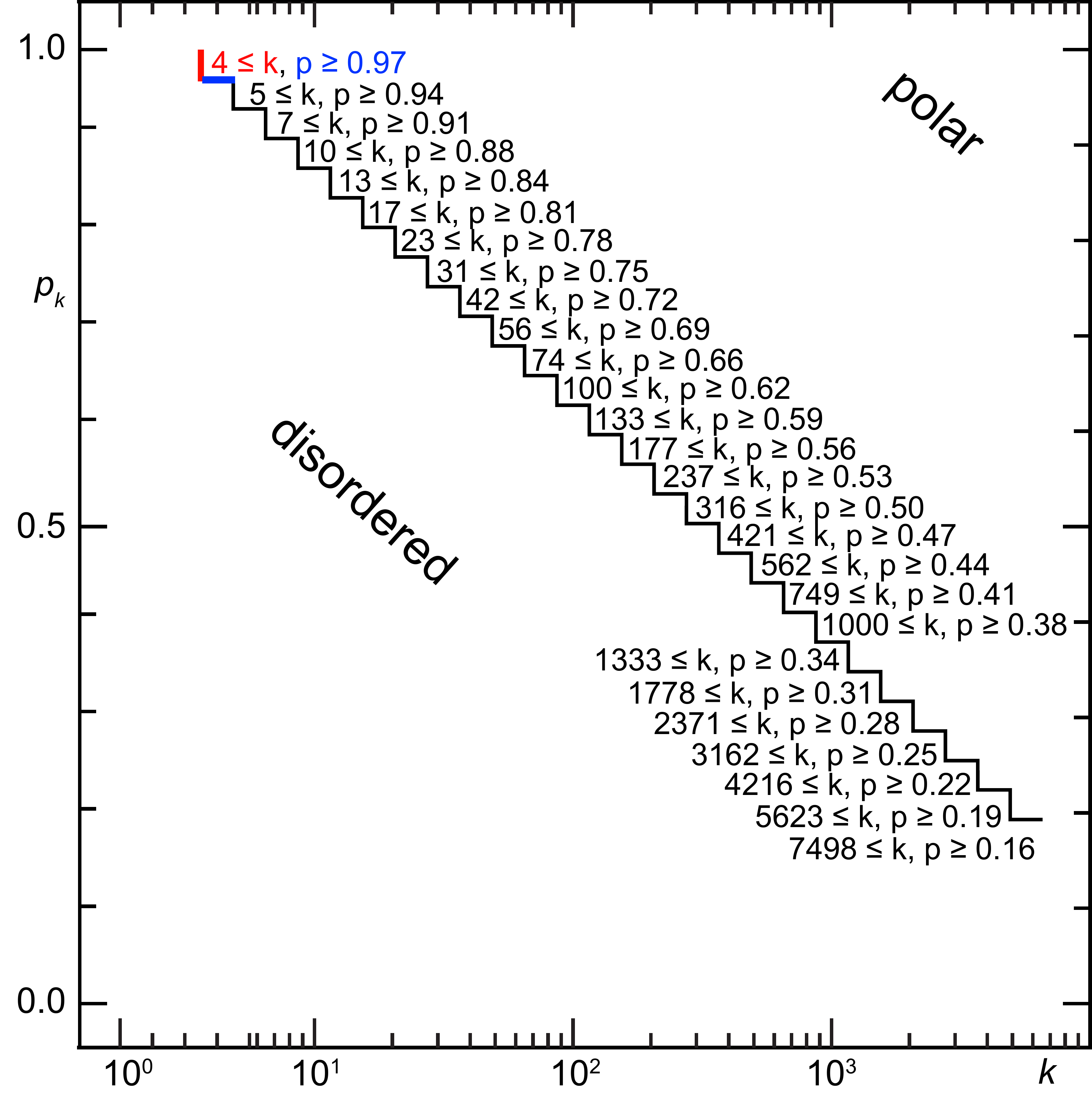}
\caption{
\textbf{Separation line between polar ordered and disordered regions.} 
Illustration of the heuristic choice for the Separation line between polar ordered and disordered regions shown in Fig.~\ref{Fig::4}(d). 
The figure shows the numerical values for the positions of the ``steps'' of the heuristic division line. 
The value next to the $k$ ($p$) denotes the position of the left (lower) boundary of a step on the $k$-axis ($p$-axis). 
The first step and the corresponding numerical values are colored for illustration.}
\label{Fig::A8}
\end{figure} 
In order to obtain the steady state particle fluxes in cluster space shown in Fig.~\ref{Fig::4}(d), we have investigated the exchange of filaments between different cluster size-order groups using agent-based simulations (WASP). 
Other than in the kinetic model, clusters in the agent-based simulations can have any degree of polar order $p_k$. 
Thus, for proper comparison, we ad hoc divided the phase space of cluster size and order (short: $k{-}p$--space) into two regions, a polar ordered and a disordered region; 
the corresponding heuristic separation line is shown in Fig.~\ref{Fig::4}(d). 
All clusters above the dividing line are defined as polar for our analysis, and all clusters below as disordered.
The line was chosen such that for a system in a disordered state, most cluster would be contained in the disordered region [cf. upper panel of Fig.~\ref{Fig::2}(c) for an example of the statistics of cluster size and order in a disordered system].
The exact numerical definition of this division line is shown in Fig.~\ref{Fig::A8}.

To measure the particle currents, we initiated a set of simulations in a polar-ordered state and recorded---in short time intervals of $\Delta t$---for each filament $j$ the temporal evolution of the size $k(t,j)$ and polar order $p(t,j)$ of the cluster to which this filament $j$ belonged to. 
With that information at hand we were able to record for any given point $(k,p)$ in the polar-ordered (disordered) region the number $\Delta n_+(k,p)$ of filaments transferred to this point from any point of the disordered (polar-ordered) region.
Likewise, $\Delta n_-(k,p)$ counts the number of filaments being transferred from this point $(k,p)$ towards any point in the disordered (polar-ordered) region.
The particle currents $J [D \leftrightarrow P_{k,p}]$  ($J [P \leftrightarrow D_{k,p}]$)
are then obtained as the difference of the counts $\Delta n_+(k,p)$ and $\Delta n_-(k,p)$ divided by the duration of the simulation. 
 For data shown in Fig.~\ref{Fig::4}(d), the simulations were performed in a steady polar-ordered state over a time period of $t \,{=}\,  50$, and we used $\Delta t \,{=}\,  0.0125$; averages were performed over $30$ statistically independent realizations. 
 Furthermore, we also determined $\int \! \mathrm{d}p \, J [D \,{\leftrightarrow}\, P_{k,p}]$ and $\int \! \mathrm{d}p \, J [P \,{\leftrightarrow}\, D_{k,p}]$
 which measure the respective currents irrespective of the specific value of cluster polar order $p$ (inset of Fig.~\ref{Fig::4}(d)). 

The difference between the current into the disordered region in the agent based simulations ($\int \! \mathrm{d}p \, J [P \,{\leftrightarrow}\, D_{k,p}]$) in the inset of Fig.~\ref{Fig::4}(d)) and in the kinetic model ($J [b {\leftrightarrow} a_k]$ in Fig.~\ref{Fig::5}(e)) is likely caused by two different factors. 
First, in our agent-based simulations, a classification of clusters into polar-ordered or disordered ones can only be done on grounds of heuristic criteria [cf. Fig.~\ref{Fig::4}(d)].
For instance, this results in more cluster sizes to be only classified as disordered, when compared with the kinetic model [cf. ``Kinetic model equations'' in the Section "Kinetic nucleation model" below.]. 
Second, in the kinetic model only disordered clusters of size $k=1$ can gain mass from polar-ordered clusters [cf. the section ``Dynamical and steady-state properties'' below], whereas in our agent-based simulations this happens also for disordered clusters larger than $1$ [cf. Fig.~\ref{Fig::4}(d)].
  
\subsection{Transition probability}\label{sup:transiton_prob}
\begin{figure}
\centering
\includegraphics[width=1.\columnwidth]{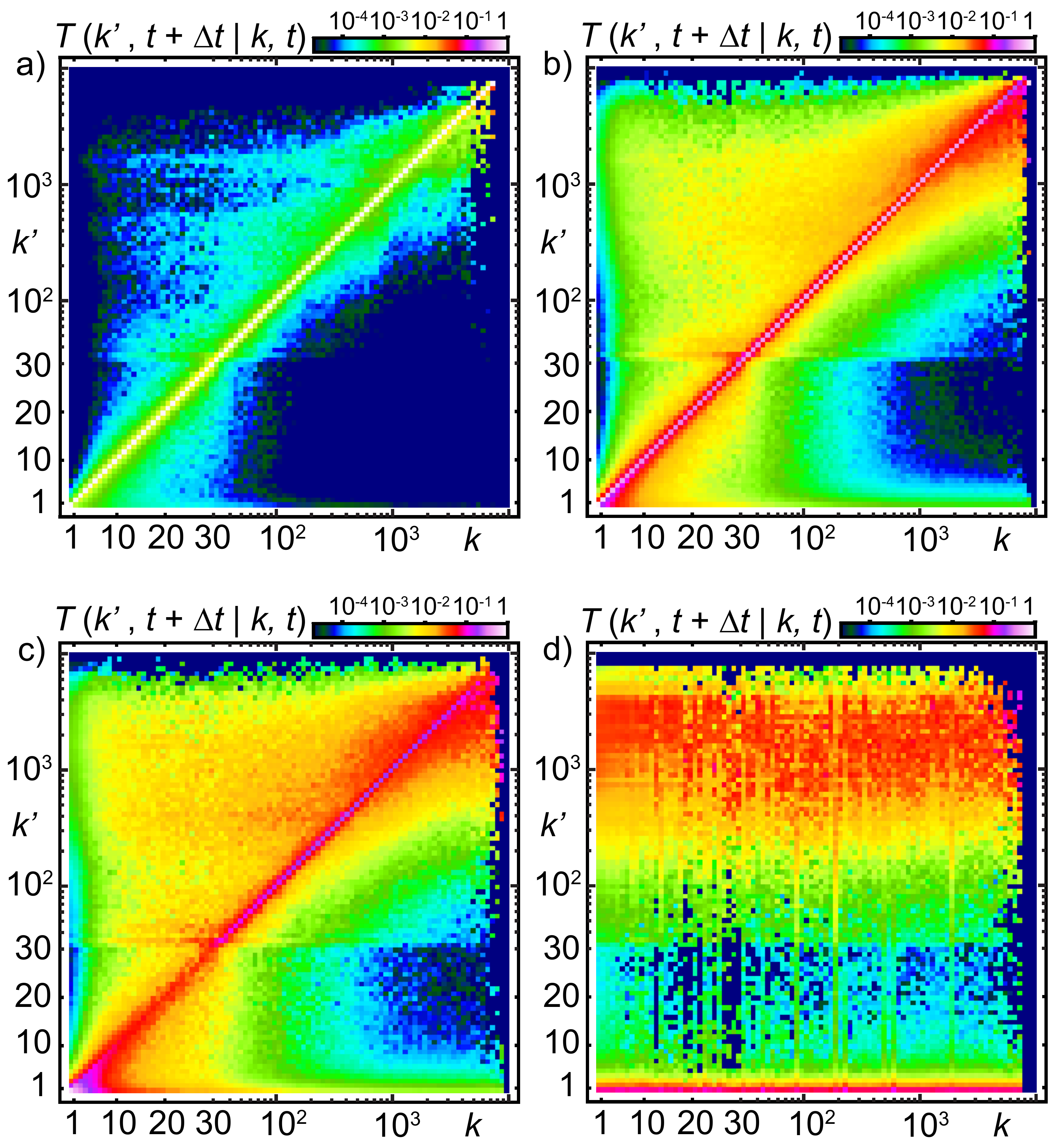}
\caption{
\textbf{Transition probabilities.} 
Matrix of transition probabilities, $T(k',t \,{+}\, \Delta t | k, t)$, in color code, for different values of the time increments $\Delta t$: a) $\Delta t$ = 0.000125, b) $\Delta t$ = 0.00625, c) $\Delta t$ =  0.025, d) $\Delta t$ = 10.0. In all panels we used $\alpha \,{=}\, 1.67$.}
\label{Fig::A9}
\end{figure}
As described in the main text, we measured the transition probabilities that a filament which is in a cluster of size  $k$ at time $t$ will be in a cluster of size $k'$ at a later time $t+\Delta t$.  

To determine these transition probabilities, $T(k',t+\Delta t | k, t)$, we used an ensemble of simulations that each was initialized in an ordered state. 
During each simulation run we recorded---in time intervals of $\Delta t$ and for each filament $j$---the size $k(t,j)$ of the cluster to which the respective filament belonged to.
We monitored for each filament all transition events from $k(t,j)$ to $k(t+ \Delta t,j)$ and collected these data in a histogram matrix $ \widetilde T^{M \times M}$ ($M$ is the number of filaments in the system). 
By normalizing its columns we obtained an approximation for the transition probabilities:
\begin{align}
	T(k',t+\Delta t | k, t) 
	\approx 
	\frac{\widetilde T(k,k')}
	     {\sum_{k'_0} \widetilde T(k,k'_0)}
	\, .
\end{align}
For the data shown in Fig.~\ref{Fig::4}(c) ($\Delta t \,{=}\, 0.0125$) we averaged the results over five simulations, which each ran for a timespan of $T \,{=}\, 50$. 

The time increment $\Delta t$ used in Fig.~\ref{Fig::4}(c) is of the same order of magnitude as the time a filament needs to travel a distance comparable to its contour length ($L/v \,{=}\, 0.0315$). 
As discussed in the main text, the precise numerical value of this increment is not important.  For comparison, Fig.~\ref{Fig::A9}(b-c) shows the matrix of transition probabilities for $\Delta t \,{=}\, 0.00625$ and $\Delta t \,{=}\, 0.025$, respectively. 
As can be inferred from this figure, they differ only on a quantitative level from Fig.~\ref{Fig::4}(c). 
For the data shown in Fig.~\ref{Fig::A9}(b-c), we averaged the results over five simulations, which each ran for a timespan of $T \,{=}\, 50$. 
The data shown in Fig.~\ref{Fig::A9} (a) and (d) ($\Delta t$ = 0.000125 and $\Delta t$ = 10) is interpreted and referenced in the discussion of the main text.
We averaged the results for $\Delta t$ = 0.000125 over five simulation runs, which each ran for $T \,{=}\, 50$. The results for $\Delta t$ = 10 were averaged over 260 simulations which also ran for $T \,{=}\, 50$.

Note that the apparent discontinuity at  $k' \,{\approx}\, 30$ is caused by changing from logarithmically arranged spacing of the binning for large cluster sizes to linear arranged spacing for small cluster sizes. 
This is necessary because clusters can only shrink or grow by integer values but a continuation of the logarithmic spacing would result in successive bin-distances becoming smaller than one. 

\section{Kinetic nucleation model}\label{sup:kinetic_nucleation_model}
\subsection{Kinetic model equations}\label{sup:kinetic_model_equations}
\begin{figure}[b]
\centering
\includegraphics[width=1.\columnwidth]{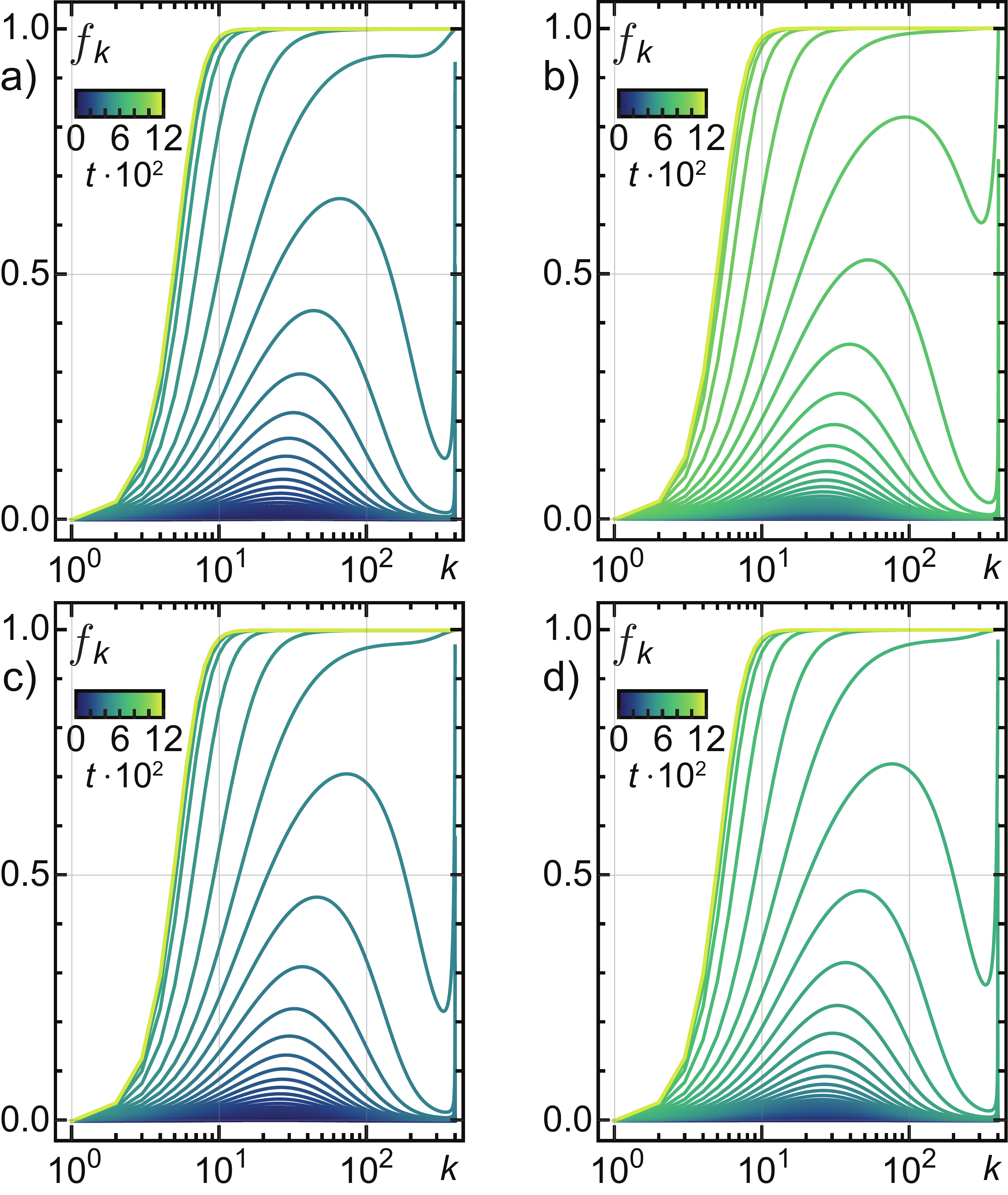}
\caption{
\textbf{Influence of $m_c$ and $v_c$ on $f_k$.} 
Time evolution of the relative fraction $f_k \,{=}\, b_k/n_k$ of ordered clusters, for different values of $m_c$ and $v_c$. 
The color gradient indicates different times as quantified by the corresponding color bar. (a) $m_c=25$ and $v_c \,{=}\, 1$ (b) $m_c \,{=}\, 200$ and $v_c \,{=}\, 1$ (c)  $m_c \,{=}\, 25$ and $v_c \,{=}\, 80$ (d) $m_c \,{=}\, 200$ and $v_c=80$.}
\label{Fig::A10}
\end{figure}
The temporal evolution of the distributions for the disordered species $a$ and the ordered species $b$ is given by:
\begin{subequations}
\begin{align} 
	\partial_t {\mathbf{a}}\label{seq:botRatesA}
	&= 
	\mathbf{F}(\mathbf{a},\mathbf{b}) 
	\, , \\
	\partial_t {\mathbf{b}}\label{seq:botRatesB}
	&=
	\mathbf{G}(\mathbf{a},\mathbf{b})
	\, ,
\end{align} 
\end{subequations}
where $\mathbf{F} \,{=}\, (F_1,F_2,...,F_M)^T$ and $\mathbf{G} \,{=}\, (G_1,G_2,...,G_M)^T$ are currents that include all possible reaction channels:
\begin{subequations}
\begin{align}
\label{eq:a}
\begin{split}
	F_1 
	&= 
	2\beta_2 \, a_2 
	+ 
	\sum_{i=3}^M \beta_i \, a_i 
	-
	\sum_{i=1}^{M-1}\alpha_{i,1} \, a_i a_1 
	\\ 
	&\quad
	+ 
	\lambda
	\Bigl(
	2 b_2 +\sum_{i=3}^M b_i
	\Bigr) 
	-
	\sum_{i=2}^{M-1} \gamma_{i,1} \, b_i a_1 
	\, , \\ 
\end{split}\\ 
\begin{split}
	F_k 
	&= 
	\beta_{k+1} \, a_{k+1} 
	- 
	\beta_k \, a_k 
	+
	\frac{1}{2}
	\sum_{i=1}^{k-1}\alpha_{i,k-i} \, a_i a_{k-i} 
	\\ 
	&\quad
	-
	\sum_{i=1}^{M-k}\alpha_{i,k} \, a_i a_k 
	-
	\sum_{i=2}^{M-k}\gamma_{i,k} \, b_i a_k 
	- 
	\omega_k \, a_k, 
\end{split}
\end{align}
\end{subequations}
and
\begin{subequations}
\begin{align} 
	G_1 
	&= 0
	\,, \\ 
	G_k
	&= 
	\lambda \, (b_{k+1} - b_k) 
	- 
	\sum_{i=2}^{M-k}\eta_{i,k} \, b_i b_k 
	\label{eq:b} \\
	&\quad 
	+ 
	\frac{1}{2}\sum_{i=2}^{k-2} \eta_{i,k-i} \, b_i b_{k-i} 
	+  
	\mu_0\left( \sum_{i=2}^{M-k} b_{i+k} - \frac{1}{2}  \sum_{i=2}^{k-2} b_k  \right) 
	\nonumber \\
	&\quad 
	+ 
	\sum_{i=2}^{k-1}\gamma_{i,k-i} \, b_i a_{k-i} 
	- 
	\sum_{i=1}^{M-k} \gamma_{k,i} \, b_k a_i + \omega_k a_k 
	\, , \nonumber 
\end{align}
\end{subequations}
with $k\in\{2,...,M\}$. 
Note that by convention, all rates are equal to zero when the indices for species $a$ are less than 1 and larger than $M$, or less than 2 and larger than $M$ for the indices of species $b$. 
It can be straightforwardly checked that these currents conserve particle mass $\sum_{k=1}^M k \, (F_k+G_k)\equiv 0$.
Please refer to the main text for the definitions and interpretations of the parameters $\beta_i$, $\alpha_{i,j}$, $\lambda_j$, $\gamma_{i,j}$, $\omega_j$, $\mu_0$ and $\eta_{i,j}$.
Note that we have assumed that clusters of size $1$ are always disordered, i.e.\ $b_1 \,{=}\, \partial_t {b}_1 \,{=}\, 0$. 
As mentioned in the main text, we fixed the parameters of the kinetic model to $M \,{=}\, 400$, $A \,{=}\, 800$, $v \,{=}\, \beta_0 \,{=}\, \lambda_0 \,{=}\, 1$, $\mu_0 \,{=}\, 0.025$, $\sigma_{aa} \,{=}\, 1.6$, $\sigma_{ab} \,{=}\, 0.2$, $\sigma_{bb} \,{=}\, 1$ and $\omega_0 \,{=}\, 10^{-4}$, if not stated otherwise.

As discussed by the authors of Ref~\cite{peruani_kinetic_2013}, $\sigma_{aa}$, $\beta_0$ and their ratio determine the shape of the distribution $a_k$ in the absence of species $b$ and exhibits a critical transition from a unimodal to a bimodal distribution. 
For our system, we took parameters such that they are always below this point to avoid structure formation in this domain.

\subsubsection{Detailed form of transformation rate}\label{sup:detailed_form_of_transformation_rate}

As noted in the main text, we have investigated how the choice of $m_c$ and $v_c$ in the expression for the transformation rate from disordered to ordered clusters ($Z(i) \,{=}\, 1/(1+e^{-(i-m_c-1)/v_c})$), influence the transition to polar order and the ordered state itself.
To this end we have performed the same kind of simulations as shown in Fig.~\ref{Fig::5}(c) but with different values for $m_c$ and $v_c$. 
As can be seen in Fig.~\ref{Fig::A10}, only the course of the transition towards polar order changes slightly. 
The stationary state, however, is identical to the one shown in Fig.~\ref{Fig::5}(c). 
This illustrates that the qualitative behaviour of the system is not sensitive to the exact choice of $m_c$ and $v_c$.  
\begin{figure}[b]
\centering
\includegraphics[width=1.\columnwidth]{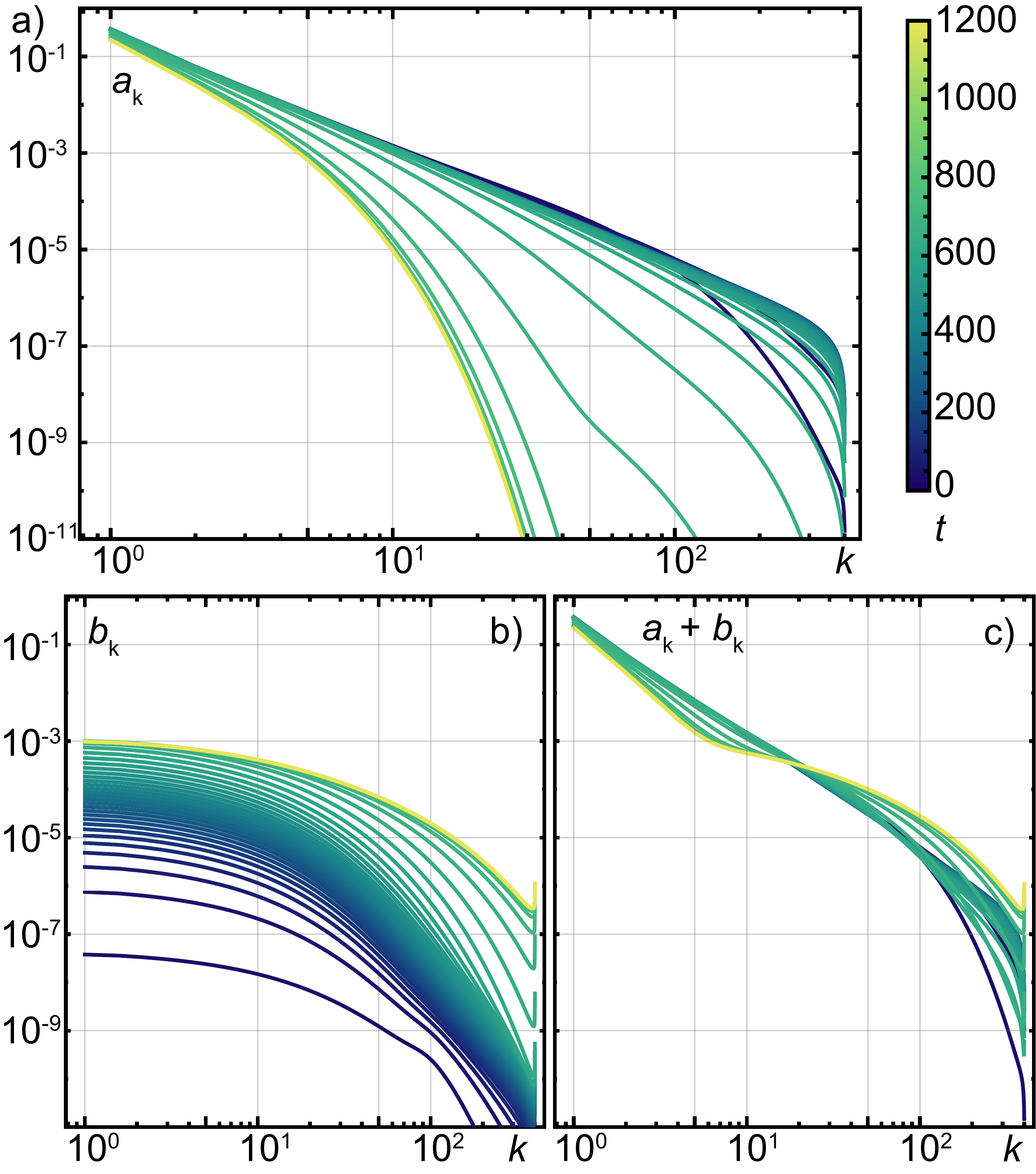}
\caption{
\textbf{Temporal evolution of the cluster distributions.} 
Temporal evolution from a disordered towards an ordered state of the single and combined cluster species distributions. 
The color gradient indicates different times as quantified by the corresponding color bar.
(a) disordered species $a_k$, (b) ordered species $b_k$, and (c) sum of both $a_k+b_k$.}
\label{Fig::A11}
\end{figure}
\subsection{Kinetic model implementation}\label{sup:kinetic_model_implementation}
We integrated Eqs.~(\ref{seq:botRatesA},~\ref{seq:botRatesB}) using a straightforward Euler scheme in C${++}$, which we found---for system sizes $M \,{\lesssim}\,  1000$---to be numerically faster than an adaptive time-step 4$^\text{th}$-order Runge-Kutta algorithm. It is furthermore simpler than implicit integration schemes, which we expect to be more stable for larger $M$.

\subsection{Dynamical and steady-state properties}\label{sup:Dynamical_and_steady_state_properties}

\subsubsection{Evolution of the cluster distributions} \label{sup:evolution_of_the_cluster_distributions}

Figure~\ref{Fig::A11} shows the temporal evolution of the polar-ordered and disordered cluster distributions, $a_k$, $b_k$, for the parameters and data shown in Fig.~\ref{Fig::5} ($\sigma_{aa} \,{=}\, 1.6$, $\sigma_{ab} \,{=}\, 0.2$).
One observes that up to intermediate times ($t\,{\approx}\, 800$) there is little change of the cluster distributions.  
Once there is a significant fraction of $b$-clusters the dynamics speeds up and their amount then increases strongly [cf. Fig.~\ref{Fig::A11}(b)]. 
This in turn leads to a substantial reduction of $a$-clusters [cf. Fig.~\ref{Fig::A11}(a)] and a corresponding change of the sum of both distributions [cf. Fig.~\ref{Fig::A11}(c)]. 

\subsubsection{Details of particle fluxes}\label{sup:details_of_particle_fluxes}

\begin{figure}[b]
\centering
\includegraphics[width=1.\columnwidth]{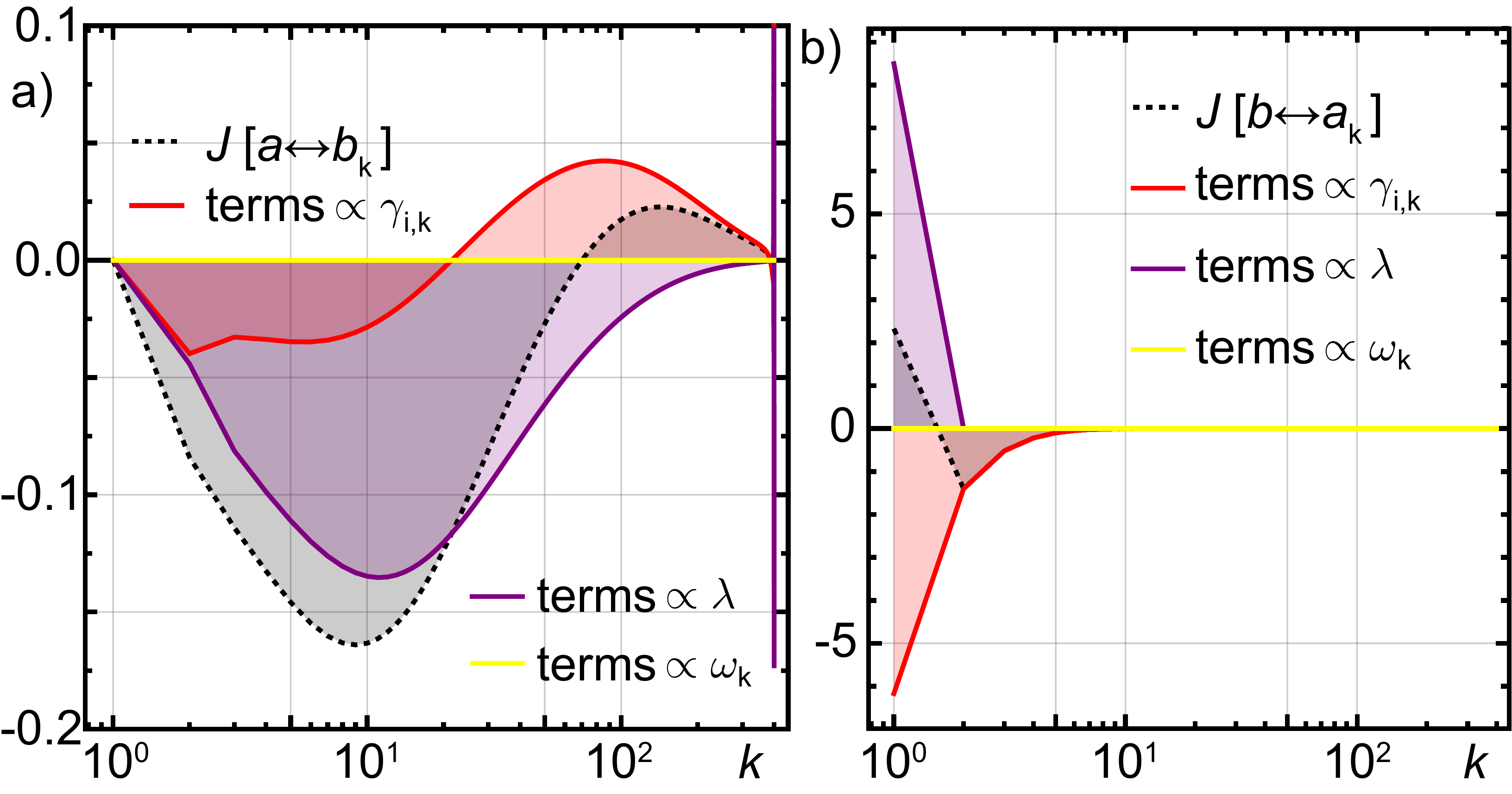}
\caption{\textbf{Contributions to inter-species fluxes.} 
Stationary inter-species particle fluxes 
$J [a {\leftrightarrow} b_k] $ (a) and $J [b {\leftrightarrow} a_k]$ (b) and individual rate contributions as a function of cluster size $k$. }
\label{Fig::A12}
\end{figure}

The inter-species fluxes of the particle mass, $J [b {\leftrightarrow} a_k]$ and $J [a {\leftrightarrow} b_k]$, as depicted in the main text in Fig.~\ref{Fig::5}(e), are obtained by setting all species-internal rates in Eq.~\eqref{seq:botRatesA} and Eq.~\eqref{seq:botRatesB} to zero. 
This leaves only inter-species contributions to $a_k$ (resulting in $J [b {\leftrightarrow} a_k]$) and inter-species contributions to $b_k$ (resulting in $J [a {\leftrightarrow} b_k]$), respectively. One obtains the following equations:
\begin{subequations}
\begin{align} 
J [b {\leftrightarrow} a_1]=\lambda(2 b_2+\sum_{i=3}^M b_i) -\sum_{i=2}^{M-1}\gamma_{i,1}b_i a_1 , \\ \label{seq:rate1}
J [b {\leftrightarrow} a_k]\overset{k>1}{=} k\cdot\left(-\sum_{i=2}^{M-k}\gamma_{i,k}b_i a_k - \omega_k a_k\right),
\end{align}
\end{subequations}
and 
\begin{align}\nonumber
&J[a {\leftrightarrow} b_k]\overset{k>1}{=} k\cdot \Bigl(\lambda (b_{k+1} - b_k) \\ \label{seq:rate2}
& + \sum_{i=2}^{k-1}\gamma_{i,k-i} b_i a_{k-i} - \sum_{i=1}^{M-k} \gamma_{k,i} b_k a_i + \omega_k a_k  \Bigr)
\, .
\end{align}

Figure~\ref{Fig::A12} illustrates the contribution of the individual currents proportional to $\lambda$, $\gamma_{i,k}$, and $\omega_k$.
This shows that species $a$ gains mass only by evaporation of single, disordered filaments from ordered clusters. 
In contrast, species $b$ gains cluster mass by coalescence of smaller ordered and disordered clusters (transferring mass to larger cluster sizes) and by transformation of disordered into ordered clusters.

\subsection{Parameter space and hysteresis}\label{sup:parameter_space_and_hysteresis}
Besides the interaction strength, the particle density is another relevant control parameter of active matter systems; e.g. in our agent-based simulations both control parameters influence the phase behaviour of the system [cf. Fig.~\ref{Fig::3}(b)].
For that reason, we investigated whether the density plays a comparable role in the kinetic model. 
To this end, we determined a bifurcation diagram as a function of $\rho_\text{kin}\,{=}\,M/A$ and $\sigma_{bb}$ (analogous to the bifurcation diagram of stationary mass fractions $\phi_b$ as a function of $\sigma_{ab}$ and $\sigma_{bb}$ [cf. Fig.~\ref{Fig::6}(c)]). 
 
\begin{figure}[]
\centering
\includegraphics[width=1.\columnwidth]{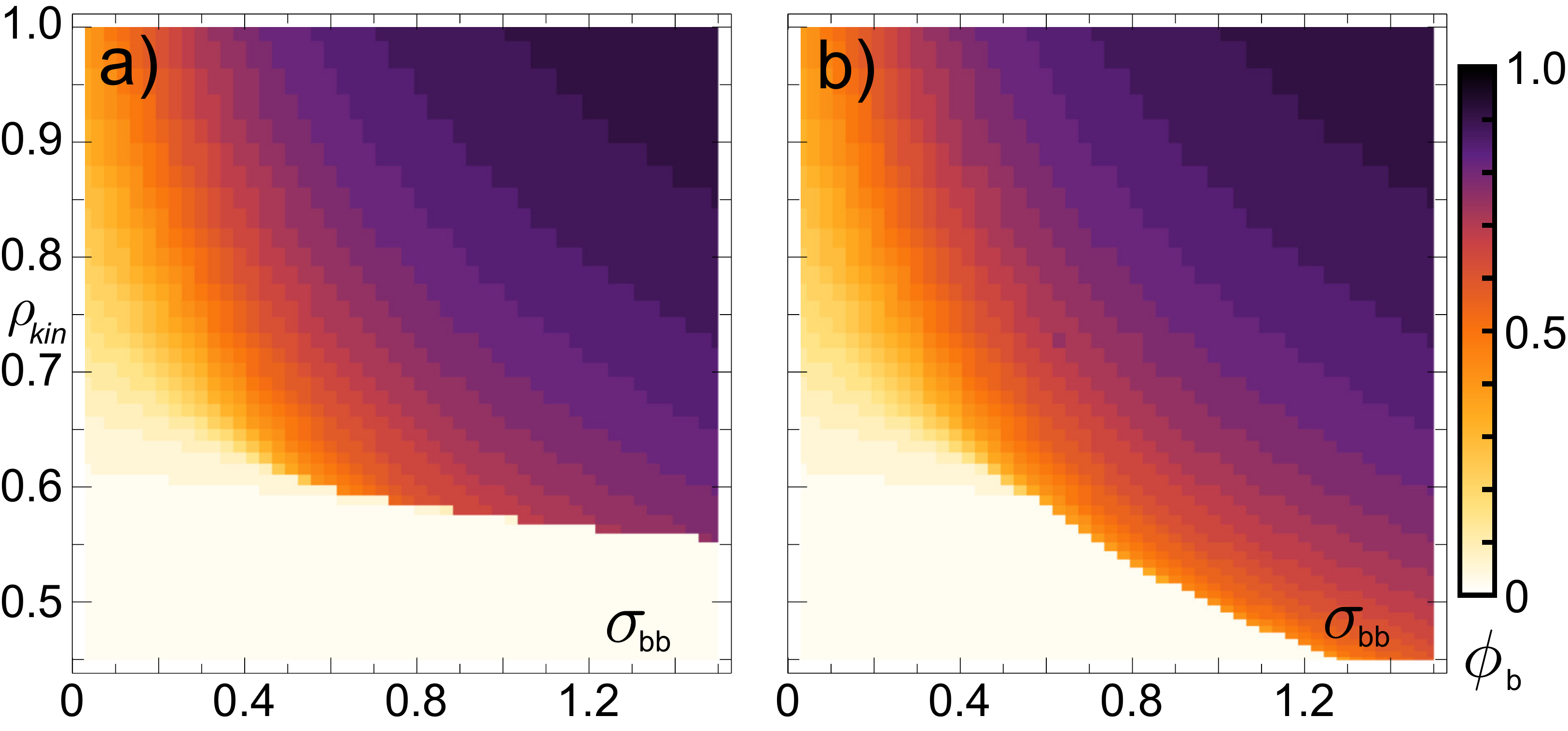}
\caption{
\textbf{$\rho_\text{kin}$-$\sigma_{bb}$ bifurcation diagram.} 
Density plot of the stationary mass fraction $\phi_b$ as a function of $\sigma_{bb}$ and the density $\rho_\text{kin}$ with different initial conditions: 
(a) started with mainly $a$ clusters present ($a_k(t{=}0) \,{=}\, \delta_{1,k}$ and $b_k(t{=}0) \,{=}\, 0$), and 
(b) started with mainly $b$ clusters present (i.e. started in a state that is similar to the stationary state in Fig.~\ref{Fig::A11}).
Parameters: $\sigma_{aa}\,{=}\,1.8$, $\sigma_{ab}\,{=}\,0.15$, $\omega_0\,{=}\,10^{-3}$.}
\label{Fig::A13}
\end{figure} 

\begin{figure}[]
\centering
\includegraphics[width=1.\columnwidth]{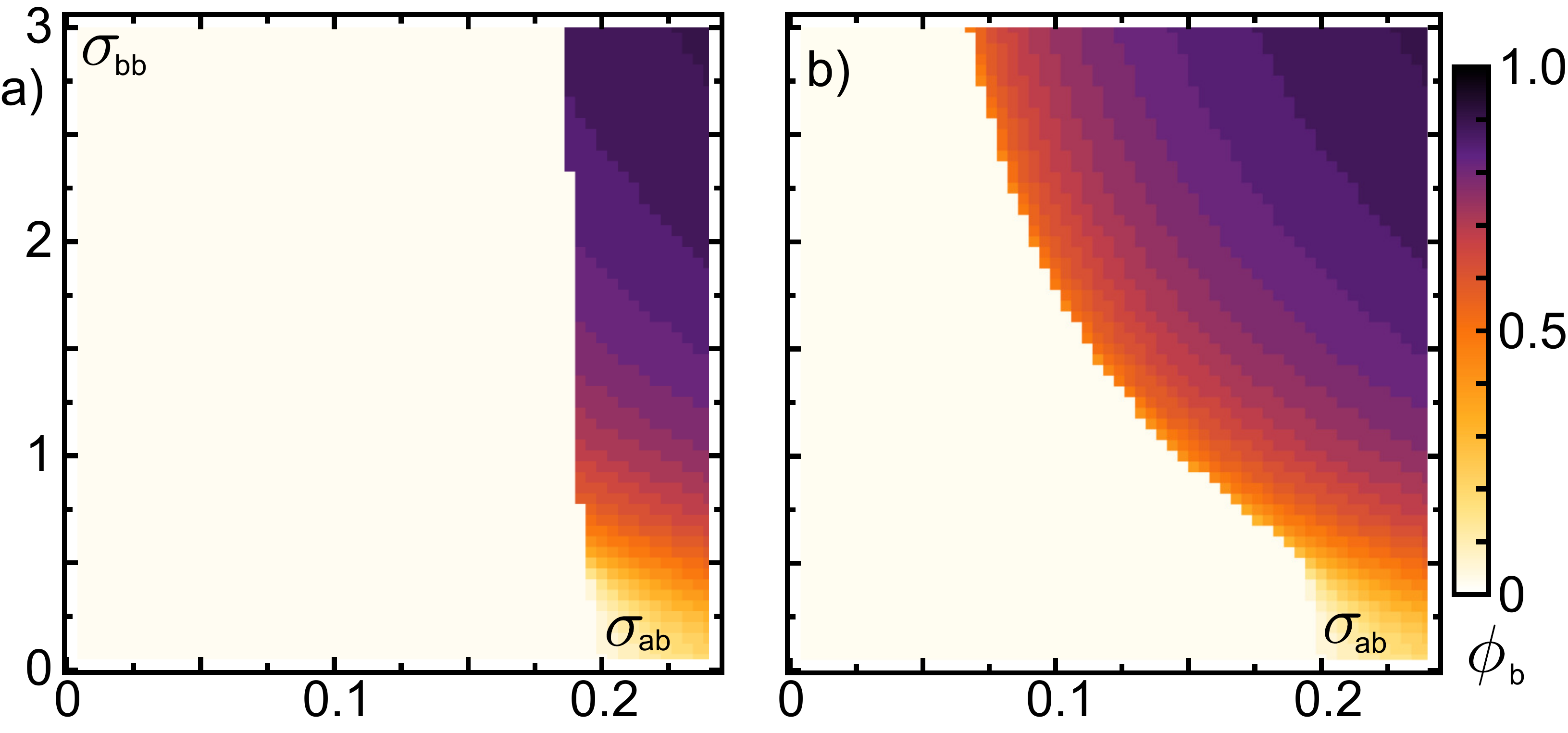}
\caption{
\textbf{$\sigma_{ab}$-$\sigma_{bb}$ bifurcation diagram.} 
Density plot of the stationary mass fraction $\phi_b$ as a function of $\sigma_{ab}$ and $\sigma_{bb}$ with different initial conditions: 
(a) started with mainly $a$ clusters present ($a_k(t{=}0) \,{=}\, \delta_{1,k}$ and $b_k(t{=}0) \,{=}\, 0$), and 
(b) started with mainly $b$ clusters present (i.e. started in a state that is similar to the stationary state in Fig.~\ref{Fig::A11}).
Parameters and data identical to Fig.~\ref{Fig::6}(b,c).}
\label{Fig::A14}
\end{figure}

Fig.~\ref{Fig::A13}(a)/(b) shows the disordered/ordered branch of the bifurcation (i.e. the stationary state of the simulations which were started in a disordered/ordered state).
Here, too, there is a bistable region between the ordered and disordered state, and, for varying the density, a discontinuity and hysteresis occurs (as it is the case for varying $\sigma_{ab}$, see Fig.~\ref{Fig::6}(b)).
In addition to the 3D-representation of the $\sigma_{ab}$-$\sigma_{bb}$ bifurcation diagram in Fig.~\ref{Fig::6}(c), and to facilitate a comparison with Fig.~\ref{Fig::A13}, Fig.~\ref{Fig::A14} shows the disordered/ordered branches of that bifurcation separately. 

\section{Supplemental movie captions} \label{sup:movie_captions} 

\textbf{Movie S1. Random nucleation and growth.}\\{\normalfont
This movie shows an agent based simulation (left side) that starts with random initial conditions. While dwelling in a disordered state, spontaneously small ordered clusters form and decay again, until, eventually, one grows large enough and triggers the formation of polar order in the system.
On the right, time-traces of the cluster polar order parameter $\Omega_p$ and the largest polar moment $S^{(1)}$ (top) and the course of the full statistics of cluster size and order $\Psi(k,p)$ (bottom) during the simulation are shown. 
(Parameters: $\alpha \,{=}\, 2$;  $\Psi(k,p)$ is calculated by a moving average over a time-window of $T \,{=}\, 0.5$ with $\Delta t \,{=}\, 0.05$. A high resolution version of this video can be found here:  
\url{https://www.theorie.physik.uni-muenchen.de/lsfrey/publication_videos/active_matter_nucleation/movie_1/}.) }\\

\textbf{Movie S2. Spontaneous build up of order.}\\{\normalfont
At high values of $\alpha$, the system does not dwell in an unordered state (in a simulation started with random initial conditions), but immediately several ordered cluster form and trigger the system to develop order without waiting time.
(Parameters: $\alpha \,{=}\, 3$. A high resolution version of this video can be found here:  
\url{https://www.theorie.physik.uni-muenchen.de/lsfrey/publication_videos/active_matter_nucleation/movie_2/}.)}\\

\textbf{Movie S3. Clusters with the largest polar moments in $k$-$p$ space. }\\{\normalfont
Course of the size $k$ and order $p_k$ of the clusters with the largest polar moments $S^{(1)}$ from 892 independent simulations. The red circles marks the average $\langle S^{(1)}\rangle$.
 Before the formation of order ($t^* \,{<}\, 0$) most clusters are located on the left of the $S_\text{crit}$-line. Only short before the systems start to develop order, the line is crossed by $\langle S^{(1)}\rangle$.
(Parameters: $\alpha \,{=}\, 1.67$. A high resolution version of this video can be found here:  \\
\url{https://www.theorie.physik.uni-muenchen.de/lsfrey/publication_videos/active_matter_nucleation/movie_3/}.)}\\

\textbf{Movie S4. Artificially inserted seeds. }\\{\normalfont
A perfectly ordered seed is placed into a system dwelling in a disordered state. Filaments that were part of the original seed are colored in magenta. The point of view is continuously shifted to the right, such that the seed stays stationary.
(Parameters: $\alpha \,{=}\, 1.67$, $4 \cdot 10^4$ filaments, $L_\text{box}\,{=}\,162.5L$, $S_\text{seed} \,{=}\, 200$. A high resolution version of this video can be found here:  \\
\url{https://www.theorie.physik.uni-muenchen.de/lsfrey/publication_videos/active_matter_nucleation/movie_4/}.)}

\cleardoublepage
\newpage

\end{document}